\documentclass[10pt]{iopart}

\usepackage{amsbsy}
\usepackage{iopams}
\usepackage{setstack}
\usepackage{amscd}
\usepackage{amsfonts}
\usepackage{amssymb}
\usepackage{graphicx}
\usepackage{amsgen}

\def\be{\begin{equation}}
\def\ee{\end{equation}}
\def\ba{\begin{eqnarray}}
\def\ea{\end{eqnarray}}
\def\bc{\begin{center}}
\def\ec{\end{center}}

\def\be{\begin{equation}}
\def\ee{\end{equation}}
\def\ba{\begin{eqnarray}}
\def\ea{\end{eqnarray}}
\def\bs{\begin{subequations}}
\def\es{\end{subequations}}
\def\e{ \rm e}
\def\dd{{\rm d}}

\def\Z#1{_{\lower2pt\hbox{$\scriptstyle#1$}}}
\def\X#1{_{\lower2pt\hbox{$\scriptscriptstyle#1$}}}
\usepackage{color}

\begin{document}

\title{On compatibility of string effective action with an accelerating universe}

\author{Ishwaree P Neupane}
\address{Department of Physics and Astronomy, University of
Canterbury, Private Bag 4800, Christchurch, New Zealand\\ and}
\address{Central Department of Physics, Tribhuvan University,
Kirtipur, Kathmandu, Nepal} \ead{ishwaree.neupane@cern.ch}

\begin{abstract}
In this paper, we fully investigate the cosmological effects of
the moduli-dependent one-loop corrections to the gravitational
couplings of the string effective action to explain the cosmic
acceleration problem in early (and/or late) universe. These
corrections comprise a Gauss-Bonnet (GB) invariant multiplied by
universal non-trivial functions of the common modulus $\sigma$ and
the dilaton $\phi$. The model exhibits several features of
cosmological interest, including the transition between
deceleration and acceleration phases. By considering some
phenomenologically motivated ansatzs for one of the scalars and/or
the scale factor (of the universe), we also construct a number of
interesting inflationary potentials. In all examples under
consideration, we find that the model leads only to a standard
inflation ($w \geq -1$) when the numerical coefficient $\delta$
associated with modulus-GB coupling is positive, while the model
can lead also to a non-standard inflation ($w<-1$), if $\delta$ is
negative. In the absence of (or trivial) coupling between the GB
term and the scalars, there is no crossing between the $w< -1$ and
$w>-1$ phases, while this is possible with non-trivial GB
couplings, even for constant dilaton phase of the standard
picture. Within our model, after a sufficient amount of e-folds of
expansion, the rolling of both fields $\phi$ and $\sigma$ can be
small. In turn, any possible violation of equivalence principle or
deviations from the standard general relativity may be small
enough to easily satisfy all astrophysical and cosmological
constraints.

\end{abstract}


\maketitle


\section{Introduction}

Dark energy, or cosmological vacuum energy, spread on every point
in space exerts gravitationally smooth and negative pressure and
hence leads to cosmic inflation, a phenomenon which is not
accounted for by ordinary matter and General Relativity (GR).
Unfortunately, at the moment the origin of dark energy (and its
associated cosmic acceleration) is arguably the murkiest question
in cosmology - and the one that, when answered, may provide
deepest insights into the origin (and the fate) of our universe,
establishing synergies between the microscopic physics and
cosmological length scales.

In recent years, the cosmic acceleration problem has been
addressed in hundreds of papers proposing various kinds of
modification of the energy-momentum tensor in the `vacuum' -- the
source of gravitationally repulsive negative pressure. Some
examples of recent interest include braneworld modifications of
Einstein's GR~\cite{DGP-2000}, scalar-tensor
gravity~\cite{Starobinsky00a}, phantom (ghost)
field~\cite{Caldwell99a}, K-essence~\cite{Picon00a} or ghost
condensate~\cite{Arkani:2003a}, $f(R)$-gravity that adds terms
proportional to inverse powers of $R$~\cite{Carroll:2003a} or
$R^n$ ($n\ge 2$) terms, or both these effects at
once~\cite{Nojirietal}, thereby modifying in a very radical manner
the Einstein's GR itself. All these proposals typically include a
more or less disguised feature, non-locality, other than classical
or quantum instabilities. A recent review by Copeland, Sami and
Tsujikawa~\cite{Samietal} summarizes the prospects and limitations
of these proposals in an elegant way.

String theory is much more ambitious and far reaching than any of
the above ideas, and it may be compatible with a de Sitter
universe (see, e.g.~\cite{KKLT}). It has been known that a
cosmological compactification of classical supergravity -- the low
energy description of string/M theory -- on compact hyperbolic
spaces~\cite{Townsend03a,Ish-2003}, or Ricci-flat twisted
spaces~\cite{Ish05-twist}, naturally leads to a period of cosmic
acceleration. Furthermore, fundamental scalar fields abundant in
such higher--dimensional gravity theories potentially provide a
natural source for gravitationally repulsive vacua, as well as a
mechanism for generating the observed density
perturbations~\cite{Mukhanov:1990me}. In this respect, the best
framework where one can hope to fully explain the dark energy
effect is perhaps string/M theory.

A strong motivation for a cosmology based on string
theory~\cite{Gasperini:1992a}, that goes beyond the standard
cosmology and incorporates Einstein's theory in a more general
framework, is that it may shed light on the nature of the
cosmological singularities arising in GR. Moreover, string theory
has the prospect of unification of all other interactions,
see~\cite{Dienes:1996a} for a review. An interesting question is
which version of string theory, if any, among the presently known
five (which are related to each other by various
dualities~\cite{Sen:1996a}), best describes the physical world. A
successful superstring theory should naturally incorporate
realistic cosmological solutions.

String theory, as a theory of quantum gravity in one {\it time}
and nine {\it space} dimensions, literally contains (or predicts)
an infinite number of scalar fields. Most of the scalars in string
theory, arising from a compactification to four-dimensions, are
structure moduli associated with the (six-dimensional) internal
geometry of space (Calabi-Yau {\it spaces} or Calabi-Yau {\it
shapes}), which do not directly couple with the spacetime
curvature tensor and hence to Einstein gravity~\cite{Witten:1996}.
It is well appreciated that, at least, one of the K\"ahler moduli
(or the modulus associated with the overall size of the internal
compactification manifold) naturally couples to (standard)
Einstein gravity via Riemann curvature
invariants~\cite{Callan85a,Narain92a}.

According to conventional wisdom superstrings involve a minimum
length scale only a few order of magnitude larger than the Planck
length, $l_{P}\sim {10}^{-33}$ cm, and, in turn, string theory
leads to a modification of Einstein's theory only at very short
distances. This idea is perfectly reasonable when we study small
fluctuations in the fixed (time-independent) background, but may
not be generally true, for the scalars such as the dilaton
($\phi$) and the modulus field ($\sigma$) and their couplings to
various curvature terms, which arise universally in any
four-dimensional heterotic superstring model, may be important in
any cosmological solution of the string effective action (see, for
example~\cite{Antoniadis93a,S-J-Rey:1996}). In this paper, we
therefore investigate as to whether the evolutions of $\phi$ and
$\sigma$, and their couplings to a Gauss-Bonnet term, can explain
the cosmic acceleration problem at early (and/or late) universe.

This paper is organized as follows. In section 2 we briefly
discuss the limitations of a standard cosmological model where
Einstein's gravity is coupled to a scalar field with a
self-interaction potential alone. In section 3 we first motivate
ourselves to a more general action which consists of, in addition
to a standard scalar potential, scalars dependent loop corrections
to the gravitational couplings of the string effective action and
write the full set of field equations for the metric, the modulus
and the dilaton. In section 4 we first write down the equations of
motion as a monotonous system of first order differential
equations, and present various analytic solutions by allowing at
most two variables of the model to take some specific forms. In
section 5, we do a similar analysis but expressing the field
equations in an alternative form, which we find more appropriate
for implementing our knowledge about the coupling functions and
the behaviour of the accelerating solutions at early or
late-times. At the end of both sections 4 and 5 we provide an
overview of the different solutions being discussed there. In
section 6, we will have a cursory look in the system where gravity
is non-minimally coupled to mater and radiation. Our conclusions
are summarized in section 7.

\section{The standard scenario}

In a standard cosmological scenario, one assumes that for a given
scalar field $\sigma$ with a self-interaction potential
$V(\sigma)$, the Einstein gravity is described by the action:
\begin{equation}\label{action1}
S= \int d^{4}{x} \left[\sqrt{-g}\left(\frac{1}{2\kappa^2}
R-\frac{\gamma}{2} (\nabla\sigma)^2 - V(\sigma)\right)\right],
\end{equation}
where $\kappa$ is the inverse Planck mass $m_{\rm Pl}^{-1}=(8\pi
G_N)^{1/2}$, $\gamma$ is the coupling constant and $\sigma$ is the
classical scalar field whose stress-energy tensor is set by the
choice of the potential. What one normally wants is a well
motivated potential for the desired evolution. This often refers
to a cosmological potential that allows at least an epoch of
cosmic inflation or acceleration/deceleration regime.

A real motivation for an effective action of the form
(\ref{action1}) arises from the fact that inflation with the
dynamics of a self-interaction potential, $V(\sigma)>0$, provides
a natural mechanism for generating the observed density
perturbations~\cite{Mukhanov:1990me}. To this end, one defines the
spacetime metric in standard Friedmann-Robertson-Walker (FRW)
form:
\begin{equation}\label{bg-metric}
ds^2=-dt^2 + a(t)^2 \sum_{i=1}^3 (dx^i)^2,
\end{equation}
where $a(t)$ is the scale factor of the universe. The Hubble
parameter that measures the expansion rate of the universe is
given by $H \equiv \dot{a}/a$, where the dot denotes a derivative
with respect to proper time $t$. One also assumes that the scalar
field $\sigma$ obeys an equation of state (EoS), with an equation
of state parameter given by
\begin{equation} w \equiv
\frac{p}{\rho}= -\frac{2}{3}\frac{\dot{H}}{H^2}-1 =
\frac{2q}{3}-\frac{1}{3},
\end{equation} where $q \equiv - a\ddot{a}/\dot{a}^2$ is the deceleration
parameter. This definition of $w$ does not depend on the number of
(scalar) fields in the model. The `equation of state' parameter
$w$ is essentially a measure of how squishy the substance is. It
is when $w<-1/3$ a gravitationally repulsive pressure exerted by
one or more scalar fields can drive the universe into an
accelerating phase ($q<0$). In particular, the cosmological
constant $\Lambda$ as a source of dark energy or vacuum energy
means that $w_\Lambda=\rho_\Lambda/p_\Lambda=-1$.

In the last two decades, many authors have studied cosmology based
on (\ref{action1}) by considering various {\it ad hoc} choice of
the scalar potential $V(\sigma)$ (the recent
reviews~\cite{Peebles:2002gy} provide an exhaustive list of
references). Written in terms of the following dimensionless
variables:
\begin{eqnarray}\label{def-noGB}
{\cal X} \equiv \kappa^2
\frac{\gamma}{2}\left(\frac{\dot{\sigma}}{H}\right)^2, \quad {\cal
Y} \equiv \kappa^2 \frac{V(\sigma)}{H^2}, \quad \varepsilon\equiv
\frac{\dot{H}}{H^2},
\end{eqnarray}
the field equations, obtained by varying the action
(\ref{action1}), satisfy the following simple
relationships~\cite{Ish:20005sd}:
\begin{equation}\label{solution1}
{\cal Y}=3+\varepsilon, \quad {\cal X}=-\varepsilon.
\end{equation} The time-evolution of the scalar field is given by
$ \sigma(t)=\kappa^{-1} \int \dd{t}[-2\dot{H}/\gamma]^{1/2}$.
Clearly, the model lacks its predictability, since $\varepsilon$
(and hence ${\cal Y}$) is arbitrary. This arbitrariness in the
potential has motivated many physicists/cosmologists to consider
models with some specific choices of the potential, e.g.
quadratic/chaotic potential $V(\sigma)= V_0 + \frac{1}{2}
m^2\sigma^2$, power-law potential $V(\sigma)= \sum m_{Pl}^{4-n}
\lambda_n\, {\sigma^n}$, inverse power-law potential
$V(\sigma)=\Lambda^4 (\sigma_0/\sigma)^n$ ($n\geq 2$), the
axion-potential $V(\sigma)= \Lambda^4 (C\pm
\cos(\sigma/\sigma_0))$, etc. This approach to the problem,
starting from Linde's model of chaotic inflation~\cite{Linde83a},
may be motivated from different particle physics models or
effective field theory. Without a complete theory of fundamental
gravity, the validity of these potentials (to explain the dark
energy problem) is not clear. The choice of possibilities for the
potential also mimics the fact that the exact nature and origin of
an inflationary potential (or dark energy) has not been
convincingly explained yet.

A common (if not conservative) argument sometimes expressed in the
literature is the following. To explain a particular value of one
new parameter (acceleration, i.e., the second derivative of the
scale factor, $\ddot{a}$), it is enough to introduce another new
parameter in the set up of the model. To this aim, one often fixes
the time dependence of the scale factor, by making a particular
{\it ans\"atz} for it, which then fixes the parameter
$\varepsilon$, and then finally find a potential, using
(\ref{def-noGB})-(\ref{solution1}), which supports this particular
$a(t)$. For instance, for the power-law expansion, $a(t)\propto
t^\alpha$, such a potential is given by
$V=V_0\,\e^{-\sqrt{2/\alpha}\,(\sigma/m_{\rm Pl})}$. In this
fashion one may find a good approximation at each energy scale,
but the corresponding solution may have little relevance when one
attempts to study a wider range of scales, including an early
inflationary epoch.

Action (\ref{action1}) is remarkably simple but not sufficiently
general as it excludes, for example, the coupling between the
field $\sigma$ (a run away K\"ahler modulus) and the Riemann
curvature invariants, as predicted correctly by some version of
string theory~\cite{Narain92a}. Also the best motivated
scalar-tensor theories which respect most of GR's symmetries
naturally include the GB type curvature invariants multiplied by
field dependent couplings. That is to say, the most general
scalar-tensor theories may involve one more parameter, the GB
integrand multiplied by a non-trivial function of $\sigma$ or
$\phi$. One may wonder whether such extra degree of freedom added
to the standard model ruins the theory, leading to a plethora of
both qualitative and analytic behaviour of cosmological solutions.
This is generally not the case; the knowledge about the $\sigma$
(and/or $\phi$)-dependent couplings may even help to construct a
realistic cosmological model with time-varying $\varepsilon$
(which otherwise remains arbitrary) without destroying the basic
characteristics of the standard model (\ref{action1}).


\section{Effective action and equations of motion}

String theory gives rise to two kinds of modifications of the
Einstein's theory. The first one is associated with the
contribution of an infinite tower of massive string modes and
leads to $\alpha^\prime$-corrections, and the second one is due to
quantum loop effects (string coupling expansion)~\cite{Callan85a}.
Therefore, if we wish a cosmological model to inherit maximum
properties of the low energy string effective action and its
associated early universe cosmology, then it may be necessary to
include an additional scalar, namely, the dilaton field $\phi$,
which plays the role of string loop expansion parameter. To this
aim, the effective action for the system may be taken to
be~\cite{Narain92a,Antoniadis93a}
\begin{eqnarray}\label{dilatonGB2}
\fl S = \int d^{4}{x}\sqrt{-g}\left[ \frac{1}{2\kappa^2}
R-\frac{\gamma}{2} (\nabla\sigma)^2-V(\sigma) -\frac{\zeta}{2}
(\nabla\phi)^2 + \frac{1}{8} [\lambda(\phi)- \delta\xi(\sigma)]
{\cal R}^2_{GB} \right],
\end{eqnarray}
where ${{\cal R}^2_{GB}} \equiv  R^2-4 R_{\mu\nu} R^{\mu\nu} +
R_{\mu\nu\rho\sigma} R^{\mu\nu\rho\sigma}$ is the Gauss-Bonnet
(GB) integrand. It is assumed that there is no potential term
corresponding to the field $\phi$. The GB term ${\cal R}^2_{GB}$
has been multiplied by non-trivial functions of $\phi$ and
$\sigma$, otherwise it would remain a topological invariant in
four dimensions. One would expect ${\cal R}^2_{GB}$ to be small in
the present epoch and that mass scales associated with the dilaton
and the modulus to be smaller than $m_{Pl}$, so that they would
not be dominating the dynamics. Only when these two conditions are
met, can the higher-derivative corrections to the action be
neglected and does string theory go over to an effective quantum
field theory~\cite{Gasperini:1992a}. Furthermore, we will consider
for most of this paper that $\gamma>0$, $\zeta>0$, so that both
$\sigma$ and $\phi$ are canonical (real) scalar fields.

It is worth noting that a GB integrand multiplied by a function of
$\phi$ is already present at the string tree-level in the next to
leading-order $\alpha^\prime$ expansion~\cite{Callan85a}, and a
non-trivial $\xi(\sigma)$ exists at the one-loop
order~\cite{Narain92a}. One also notes that the (semi)classical
tests of Newtonian gravity typically deal with small fluctuations
in the fixed (time-independent) background and thus are unaffected
by the GB type modification of Einstein's theory. Another beauty
of the action (\ref{dilatonGB2}) is that the equations of motion
for the modulus-dilaton-graviton system form a set differential
equations with no more than second derivatives of the fields or
the metric, which can be solved exactly with some specific choice
of the (coupling) parameters, or {\it a prior} knowledge about the
evolution of one of the scalars, and/or the scale factor, $a(t)$.

As compared to the four-dimensional perturbative string effective
action obtained from a compactification of 10d heterotic
superstring on a symmetric orbifold~\cite{Narain92a}, action
(\ref{dilatonGB2}) lacks the terms proportional to
$R\tilde{R}=\eta^{\mu\nu\rho\lambda} R_{\mu\nu}\,^{\sigma\tau}
R_{\rho\lambda\sigma\tau}$ (where $\eta^{\mu\nu\rho\lambda}\equiv
\epsilon^{\mu\nu\rho\lambda}/\sqrt{-g}$ and
$\epsilon^{\mu\nu\rho\lambda}$ a totally anti-symmetric tensor)
which however give a trivial contribution in the background
(\ref{bg-metric}). The numerical coefficient $\delta$ typically
depends on the massless spectrum of every particular model;
in~\cite{Narain92a}, it was found to be associated with the
four-dimensional trace anomaly of the $N=2$ supersymmetry of the
theory. Here we absorb this coefficient into the coupling function
$\xi(\sigma)$.

An often asked question is the following: what are the choices for
$V(\sigma)$, $\xi(\sigma)$ and $\lambda(\phi)$)? And what could
motivate these choices? One can take a different point of view: it
may be more productive to know what form of the potential (or the
scalar-GB couplings) offers the best resolution to the dilemma
posed by the current cosmic acceleration or the dark energy
problem. In some perturbative string amplitude computations,
$V(\sigma)$ appeared to be absent~\cite{Narain92a}, implying that
it is a phenomenologically-motivated term. Whatever the origin of
the potential, $V(\sigma)$, in (\ref{dilatonGB2}) may be, we
believe that the $\sigma$-dependence of the potential is quite
universal. This was also the case revealed recently from type IIB
string theory compactified on a (deformed) conifold~\cite{KKLT};
this study also includes some supersymmetry breaking
non-perturbative effects, mainly those coming from different
branes and fluxes present in the extra dimensions. Typically, the
moduli potentials are generated by fluxes, so $V(\sigma)$ in
(\ref{dilatonGB2}) may take into account a contribution from
fluxes.

Recently, \cite{Nojiri05b} examined a cosmological solution for
the system~(\ref{dilatonGB2}) with the simplest choice of the
scalar potential and the modulus-GB coupling, namely
$V(\sigma)=V_0 {\rm e}^{-2\sigma/\sigma_0}, \quad
\xi(\sigma)=\xi_0 {\rm e}^{2\sigma/\sigma_0}$,
deleting the $\phi$-dependent terms. Even this simple system led
to plethora of qualitative behaviors of cosmological solutions,
among which one may accommodate lots of data, including the dark
energy's `equation of state' parameter, $w$. However, a solution
of the type $a(t)\propto t^\alpha$ with $\alpha>0$, or
$a(t)\propto (t_\infty-t)^{\alpha}$ with $\alpha<0$, i.e. a
power-law expansion for which $\varepsilon=$ const, does not
support the transition from acceleration to deceleration as well
as the crossing between quintessence ($w>-1$) to phantom ($w<-1$)
phases. A non-constant $\varepsilon$ is generally required for
achieving a natural exit from inflation in the early universe
cosmology as well as for explaining a transition from deceleration
to acceleration in the expansion of the (late) universe, which has
been viewed as a recent phenomenon~\cite{Bennett03a}. Thus we
shall normally consider a situation in which $\varepsilon$ (and
hence the EoS parameter, $w\equiv -2\varepsilon/3-1$) dynamically
changes as it happens in inflationary cosmology.

A study has been recently made in~\cite{follow-up} by considering
a variant of the action~(\ref{dilatonGB2}), which includes
additional $\phi$-dependent higher curvature terms and also a
non-trivial coupling between $\phi$ and the Einstein-Hilbert term,
i.e. in the presence of a single scalar field (a dilaton or
compactification modulus). The method authors used there to find
the solutions was mainly to identify analytically some asymptotic
solutions and then try to join them numerically. A similar
approach was taken previously in the paper~\cite{Antoniadis93a},
for the action (\ref{dilatonGB2}) with $V(\sigma)=0$
(\cite{Easther:1996a,Kanti:1998jd} provided further
generalizations with a non-zero spatial curvature). In this work,
we examine the system analytically, by writing the field equations
as a monotonous system of first (and second) order differential
equations. In the background (\ref{bg-metric}), the equations of
motion derived from the action~(\ref{dilatonGB2}) are given by
(see, e.g., \cite{Antoniadis93a,Ish00a})
\begin{eqnarray}
&& 0= -\frac{3}{\kappa^2} H^2 +\frac{\gamma}{2}\dot{\sigma}^2
+\frac{\zeta}{2}\dot{\phi}^2+V(\sigma)
-3 H^3 \dot{f}, \label{GB1}\\
&&0= \frac{1}{\kappa^2} (2\dot{H}+3H^2)+ \frac{\gamma}{2}
\dot{\sigma}^2+\frac{\zeta}{2}\dot{\phi}^2 -V(\sigma)+ H^2
\ddot{f} + 2 H \dot{H}\dot{f}+2 H^3\dot{f},~~
\label{GB2}\\
&& 0= \gamma(\ddot{\sigma}+3H\dot{\sigma}) +\frac{d
V(\sigma)}{d\sigma}+\frac{1}{8}\,\frac{d\xi(\sigma)}{d\sigma}\,{\cal
R}^2_{GB},
\label{GB3}\\
&&0= \zeta(\ddot{\phi}+3H\dot{\phi})-\frac{1}{8}\,\frac{d
\lambda(\phi)}{d\phi}\,{\cal R}^2_{GB}, \label{GB4}
\end{eqnarray}
where $f\equiv \lambda(\phi)-\xi(\sigma)$. Note that $\phi=$ const
is a solution of the last equation above only if $\lambda(\phi)=$
const, implying a trivial coupling between $\phi$ and the GB term.
Because of the Bianchi identity, one of the equations is redundant
and may be discarded, as it will be trivially satisfied by a
solution obtained by solving three equations only: equations
(\ref{GB1}) and (\ref{GB2}) are not functionally independent. More
precisely, the linear combination $ H \times (7)  + H \times (8)
 -\frac{1}{3}\, {\dot\sigma} \times (9) - \frac{1}{3}\,
{\dot\phi} \times (10)$ yields the time derivative of (7). One may
drop, for instance, (8), so that (7), (9) and (10) form a set of
independent equations of motion.

In~\cite{Ish:20005sd}, we analysed the model by treating $\phi$ as
a constant, which then leaves only $\sigma$ as a dynamical field.
This analysis was partly motivated from an earlier observation by
Antoniadis et al.~\cite{Antoniadis93a}, in the $V(\sigma)=0$ case,
that the behaviour of solutions depends crucially on the form of
$\sigma$-dependent string one-loop corrections, while the
$\phi$-dependent contribution is negligible and may be ignored.
This has got another clear motivation: in string theory context,
fixing of the volume (or K\"ahler) moduli, is only an approximate
idea (see, e.g.\cite{KKLT,Kachru02a}).

We are not interested in the situation where the contributions
coming from the GB term dominates the other terms in the effective
action, which is consistent with our approximation of neglecting
higher derivative terms, like $(\nabla\sigma)^4$ and also higher
curvature terms. For all solutions to be discussed below, the
squared of the Hubble parameter, $H^2$, will decay faster than the
couplings $\lambda(\phi)$ and $\xi(\sigma)$ grow
with time. 
Thus the moduli/dilaton dependent two (and higher) loop
corrections to higher derivative terms or terms higher than
quadratic in the Riemann tensor, which give at least sixth powers
of $H$, are small and may be neglected. Nevertheless, it is
interesting to note that a four-dimensional string effective
action without a Gauss-Bonnet term, but including terms up to
quartic in $\sigma$, $(\nabla\sigma)^4$, give rise to solutions
supporting inflation and/or dark energy universe. There also exist
models where an accelerated expansion can be realized without a
field potential; some examples are the K-inflation and dilatonic
ghost condensate~\cite{Picon-et-al}.

We study here a model without matter couplings; only the matters
that are present in a fundamental theory such as a string/M theory
arise from the decay of the fields, like $\sigma$ and $\phi$. In
the last section, however, we will have a cursory look in the
system where $\phi$ is coupled (non-)minimally to ordinary matter
and radiation.

\section{First-order system}

Let us define the following dimensionless variables:
\begin{eqnarray}\label{variables1}
{\cal X} & \equiv& \kappa^2
\frac{\gamma}{2}\left(\frac{\dot{\sigma}}{H}\right)^2, \qquad
{\cal Y} \equiv \kappa^2 \frac{V(\sigma)}{H^2}, \qquad  {\cal
W}\equiv
\kappa^2\frac{\zeta}{2}\left(\frac{\dot{\phi}}{H}\right)^2,
\nonumber \\
{\cal U} &\equiv& \kappa^2 \dot\sigma {H}
\frac{\dd\xi(\sigma)}{d\sigma},
 \qquad
{\cal V} \equiv \kappa^2 \dot\phi {H}
\frac{\dd\lambda(\phi)}{d\phi},\qquad \varepsilon\equiv
\frac{\dot{H}}{H^2}.
\end{eqnarray}
In terms of these variables, equations (\ref{GB1})-(\ref{GB4})
take the following form (see the Appendix for details)
\begin{eqnarray}
0&=& -3+{\cal X}+{\cal Y}+{\cal W}+3{\cal U} -3{\cal V}, \label{eq1a}\\
0 &=& 2\varepsilon+3 + {\cal X}- {\cal Y}+{\cal W} + {\cal
V}^\prime-{\cal U}^\prime
 +(\varepsilon+2) ({\cal V}-{\cal U}),\label{eq2a} \\
0&=& {\cal X}^\prime+2{\cal X}(\varepsilon+3)+{\cal Y}^\prime
+2{\cal Y} \varepsilon
+3{\cal U} (\varepsilon +1 ),\label{eq3a} \\
0&=& {\cal W}^\prime+ 2{\cal W} (\varepsilon+3) -3{\cal V}
(\varepsilon +1 ),\label{eq4a}
\end{eqnarray}
where the prime denotes a derivative with respect to $N$, where $
N=\ln (a(t)/a_0)=\int H \dd t$ is the number of e-folds. We also
note that $\dot{\sigma}/H=\sigma^\prime$ and
$\dot{\phi}/H=\phi^\prime$.

In our discussion we normally assume that the scale factor of the
universe before inflation is $a_0$, so initially, $N\equiv \ln
(a(t)/a_0)=0$. At late times, one has $N\gg 0$, since $a(t)\gg
a_0$. However, this assumption may just be reversed, especially,
if one wishes to apply the model to study a late-time cosmology,
where one normalizes the scale factor such that its present value
is $a_0$, which then implies that $N< 0$ in the past, since $a(t)<
a_0$.

In a known example of heterotic string
compactification~\cite{Narain92a,Antoniadis93a}, one defines, to
leading order in string loop expansions, $\lambda(\phi)=\lambda_0
\e^{\phi}$ and $\xi(\sigma)=\ln 2-
\frac{\pi}{3}\,\e^{\sigma}+\sigma+4\sum_{n=1}^\infty \ln
(1-\e^{-2n\pi \e^{\sigma}})$; the latter implies that
$\dd\xi/\dd\sigma\sim - {\rm sgn}(\sigma) \frac{2\pi}{3}
\sinh(\sigma)\le 0$ and hence ${\cal U}>0$ ($<0$) for
$\dot{\sigma} H<0$ ($>0$). For other choice of $\lambda(\phi)$,
$\xi(\sigma)$ can be different. For this reason, as well as for
generality, we would prefer to keep both $\lambda(\phi)$ and
$\xi(\sigma)$ arbitrary. At any rate, the scalar GB couplings of
the above forms may be realized as special limits of a more
general class of solutions discussed in this paper.

First we note that the system of equations
(\ref{eq1a})-(\ref{eq4a}) may be expressed broadly into the
following two categories (branches):
\begin{eqnarray}\label{GB-first-branch}
&& {\cal Y} = 3 +\varepsilon -\frac{1}{2} {\cal U}^\prime-
\frac{1}{2} (5+\varepsilon) {\cal U} +\frac{1}{3}(1+\varepsilon)
{\cal W},
\nonumber \\
&& {\cal X} = -\varepsilon + \frac{1}{2} {\cal
U}^\prime-\frac{1}{2}
(1-\varepsilon){\cal U}  + \frac{1}{3}(2-\varepsilon) {\cal W},\nonumber\\
 &&
{\cal W}= \e^{-4( N- N_0) }, \qquad \varepsilon=\varepsilon( N),
\qquad {\cal U}={\cal U}( N),
\end{eqnarray}
for ${\cal V}=\frac{2}{3} {\cal W}$, where $N_0$ is an integration
constant, and
\begin{eqnarray}\label{GB-second-branch}
{\cal Y} &=& \frac{{\cal V}^\prime-{\cal U}^\prime}{2}+{\cal
V}-{\cal U}+ \frac{({\cal W}^\prime+6{\cal V})({\cal U}-{\cal
V}-2)}{2(2{\cal
W}-3{\cal V})},\nonumber\\
{\cal X} &=& \frac{{\cal U}^\prime-{\cal V}^\prime}{2}-
\frac{({\cal W}^\prime+8{\cal W}-6{\cal V})({\cal U}-{\cal
V}-2)}{2(2{\cal W}-3{\cal V})}-{\cal W} -1,\nonumber\\
\varepsilon &=& \frac{-{\cal W}^\prime-6{\cal W}+3{\cal V}}{2{\cal
W}-3{\cal V}},
\end{eqnarray}
for ${\cal V}\neq \frac{2}{3}{\cal W}$. In the first case, for
$N\gg N_0$, the time variation of $\phi$ is exponentially
suppressed, $\phi^\prime\simeq 0$.

Below we present a number of interesting cosmological solutions
(as well as construct inflationary potentials), by considering
some {\it a prior} information about the evolution of one of the
scalar fields and/or the scale factor. Our discussions also
generalize past investigations in the literature where comparison
is possible.

\subsection{Deleting $\sigma$-dependent terms}

If we drop the $\sigma$-dependent terms, then we find
\begin{equation}\label{no-sigma-terms}
\fl \zeta {\phi^\prime}^{\,2}- 6\kappa^2 \phi^\prime H^2
\lambda_{,\,\phi}-\frac{6}{\kappa^2}=0,\qquad
\zeta\kappa^2\phi^\prime \phi^{\prime\prime}+\frac{\zeta}{2}\,
\kappa^2(5+\varepsilon){\phi^\prime}^{\,2}+3\varepsilon+3=0
\end{equation}
as the set of two independent equations. It is not difficult to
see that $\phi^\prime$ which decreases faster than $1/a^2$ can
lead to an accelerated expansion, that is, $\varepsilon>-1$. Note
that in general $N$ depends on the field $\phi$, and its first
time derivative, $\dot{\phi}$ (or $\phi^\prime$). However, if the
slow roll-condition $\zeta\phi^\prime\simeq \lambda_{\,\phi} H^2
(1+\varepsilon)$ is satisfied at a particular epoch, then in that
epoch $N$ primarily depends on the field value. The simplest
ans\"atz is $\phi\equiv \phi_0 N+$ const. Then we find
\begin{equation}
H(\phi) = H\Z0 \,\e^{-\,p(\phi/\phi_0)},\qquad \lambda(\phi) =
\lambda_0\,\e^{2p(\phi/\phi_0)}+\lambda_1,
\end{equation}
where $p =(6+5\zeta\kappa^2\phi_0^2)/(6+\zeta\kappa^2\phi_0^2)$.
For a canonical $\phi$ ($\zeta>0$), this solution does not give a
cosmic acceleration.

For instance, in an inflationary model motivated by brane worlds,
one may consider a modified ans\"atz: $\Delta N\equiv (1/\alpha)
\ln(\phi/\phi_0) \equiv (1/\alpha)\ln\tilde{\phi}$; here $\phi_0$
may characterize the minimum separation between two moving branes.
Hence
\begin{equation}
H(\phi)  = H\Z0\, \tilde{\phi}^{-1/\alpha} \left(6+\mu
\tilde{\phi}\right)^{-1-4/\alpha}, \qquad \lambda_{,\,\phi} =
\frac{\mu \tilde{\phi}-6}{6\kappa^2\alpha\phi H^2},
\end{equation} where $\mu\equiv \zeta\kappa^2\phi_0^2\alpha^2$.
The universe accelerates for $\alpha<-4$. Of course, one may solve
the system of equations~(\ref{no-sigma-terms}) by making an even
more complicated ans\"atz, such as $\phi=\phi_0 \tanh(\alpha
\Delta N)$, which may motivated, for example, by some domain wall
solutions. The two examples that we discussed above are of simpler
kind.

\subsection{Deleting $\phi$-dependent terms}

If we drop all $\phi$-dependent terms or simply consider a
constant dilaton background, then we find
\begin{equation}\label{only-sigma-eqs}
{\cal X}+ {\cal Y} + 3 U = 3, \qquad U^\prime-(1-\varepsilon)
U-2(X+\varepsilon)=0.
\end{equation}
First, as is often the case in most inflationary models,
satisfying slow-roll conditions, we shall assume that $N\equiv
\sigma/\sigma_0+$ const. Then there exist two classes of
solutions: the first class of solutions is characterized by
\begin{equation}\label{Nojiri-type}
\fl \xi(\sigma)= \xi\Z0\,\e^{\beta\sigma/\sigma\Z0}+\xi_1, \qquad
V(\sigma) =\frac{1-\delta}{\kappa^2}\,H(\sigma)^2
=\frac{2\left(1-\delta\right)}{3\kappa^4\, \xi^\prime} \equiv
V\Z0\,\e^{-\beta\sigma/\sigma\Z0},
\end{equation}
where $\beta\equiv 1+3\delta$ and $\delta\equiv
\gamma\kappa^2\sigma\Z0^2/2$. These are nothing but the GB
coupling and the scalar potential considered, for instance,
in~\cite{Nojiri05b}; such terms may be generated also using the
formalism discussed in~\cite{Nojiri:2006je}). The second class of
solution may be characterized by the potential
\begin{equation}
V(\sigma)=(3-3\kappa^2 H^2\xi^\prime-\delta) H^2,
\end{equation}
where the Hubble parameter is a solution of the differential
equation:
\begin{equation}
\frac{H^\prime}{H}=\frac{\kappa^2
H^2(\xi^{\prime\prime}-\xi^\prime)-2\delta}{2-3\kappa^2
H^2\xi^\prime}.
\end{equation}
Especially, for the power-law expansion, $H\equiv
H_0\,\e^{\varepsilon_0 N}$, or equivalently $a(t)\propto (c_0
t+t_1)^{p}$, we find
\begin{equation}
\xi(\sigma)=\xi_0 \e^{2 p (\sigma/\sigma_0)} +\xi_1\,\e^{(1+3
p)(\sigma/\sigma_0)}+\xi_2,
\end{equation}
where $\xi_0\equiv p(1-\delta p)/[(1+p)\kappa^2 H_0^2]$, and
$\xi_1$ and $\xi_2$ are arbitrary constants. Unlike result
(\ref{Nojiri-type}), the GB coupling is now a sum of two
exponential terms.

In fact, alternatively, one can solve the equations
(\ref{only-sigma-eqs}) by making reasonable {\it ans\"atzes} for
two of the variables; in particular, for ${\cal X}\equiv x_0
\,\e^{2\alpha_1 N}$ and ${\cal U}\equiv u_0\,\e^{\alpha_2 N}$, we
find $V(\sigma)=3 m_{Pl}^2 H^2 (1-\lambda_0 \sigma^2-\lambda_1
\sigma^{\alpha_2/\alpha_1})$, where $\lambda_0$ and $\lambda_1$
take the sign of $x_0$ and $u_0$, respectively. Similarly, with
$\alpha_1\simeq 0$, we find $V(\sigma)=m_{Pl}^2 H^2
(V_0+V_1\,\e^{-\alpha_2\sqrt{\gamma/x_0}\,\sigma})$. The
cosmological applications of these potentials will be discussed
elsewhere.

\subsection{Late-time cosmology: two scalar case}

Consider that the field $\sigma$ is rolling slowly such that
$\kappa |\sigma^\prime|\equiv \sqrt{2\sigma_0/\gamma}$ and $\kappa
H^{-1}\equiv \sqrt{y_0}\, V^{-1/2}$, where $\sigma_0$ is a (small)
constant and $y_0>0$. These are good approximations for late-time
cosmology. Then from (\ref{GB-first-branch}) we find
\begin{equation}\label{late-sol1}
\fl \kappa \phi^\prime = \pm
\sqrt{\frac{2}{\zeta}}\,\e^{-\,2(N-N_0)}, \quad \kappa
|\xi_{,\,\sigma}|=\frac{1}{H^2}\sqrt{\frac{\gamma}{2\sigma_0}}\,
\frac{A_-}{3}, \quad \varepsilon =-1+\frac{2(y_0-2\sigma_0)}{A_+},
\end{equation}
where $A\Z\pm \equiv 3+
\frac{\zeta}{2}\,\kappa^2{\phi^\prime}^{\,2} \pm (\sigma\Z0+y\Z0)$
and
\begin{equation}
H \propto \e^{(m-1) N}\, {A_+}^{m/4},
\end{equation}
where $m \equiv 2(y\Z0-2\sigma\Z0)/(\sigma\Z0+y\Z0+3)$. A cosmic
acceleration occurs for $m>0$, which is equivalent to the
condition $V(\sigma)> 2 K(\sigma)$ in the standard model, i.e.,
without the GB coupling. A phantom-like equation of state, namely
$w<-1$ (or $\varepsilon>0$) may be obtained by satisfying
$\zeta\kappa^2 {\phi^\prime}^{\,2}< 2(y_0-5\sigma_0-3)$, or $m>1$
if $\phi$ is constant.

One notes that $\phi^\prime=-\,\frac{\phi}{2}+ c$, where $c$ is an
arbitrary constant. The scalar potential may be given by
\begin{equation}
V(\sigma)= \e^{\kappa \beta_1 \sigma}\,
\left(\Lambda_0+\Lambda_1\,\e^{-\,\kappa\beta_2\sigma}\right)^{m/4},
\end{equation}
where $\beta_1\equiv (m-1)\sqrt{\frac{\gamma}{2\sigma_0}}$ and
$\beta_2\equiv \sqrt{\frac{8\gamma}{\sigma_0}}$. Equivalently,
\begin{equation}
V(\sigma)\to V(\sigma,\phi) =\Lambda_0\,\e^{\kappa \beta_1
\sigma}\, \left[\kappa^2\left(\phi^2-4 c \phi\right)\right]^{m/4},
\end{equation}
up to a shift in $\phi$. A scalar potential exponential in
$\sigma$ (and multiplied by a non-trivial function of the dilaton
field $\phi$) may be motivated in string theory, via a study of
gaugino condensation; in this context, $\sigma$ may be viewed as
one of the K\"ahler moduli (see, e.g.~\cite{Font:1990,KKLT}).

\begin{figure}
\bc
\includegraphics[width=3.2in]{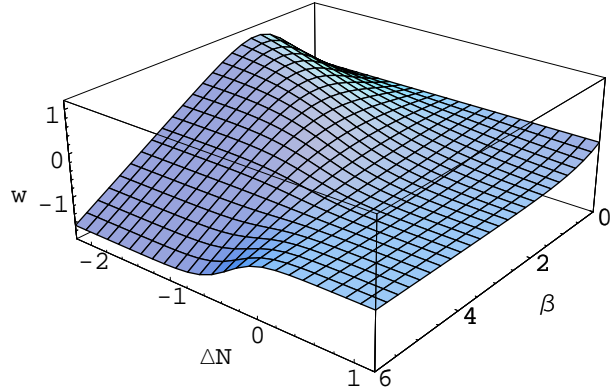}
\ec \caption{\label{w-vs-eta-latetime} A non-constant dilaton
case. The parameters are chosen as $y_0=3$, $\sigma_0=3/5$ (cf
equation~(\ref{non-const-dilaton1})) and $\phi^\prime \propto
\e^{-\beta\Delta N/2}$. There exists a solution crossing the
barrier $w=-1$, only if $\beta\gtrsim 5$. For $\beta<5$, the
universe passes smoothly from a decelerating phase ($w>-1/3$) to
an accelerating phase ($w<-1/3$), but it does not cross the
barrier $w=-1$. }
\end{figure}

We can manipulate the system of equations (\ref{GB-second-branch})
so as to obtain the expression for $\varepsilon$:
\begin{equation}\label{non-const-dilaton1}
\varepsilon = -1+\frac{2 (y_0-2\sigma_0)-{\cal W}^\prime-4{\cal
W}}{\sigma_0+ y_0+3+{\cal W}},
\end{equation}
where ${\cal W}=\frac{\zeta}{2}\kappa^2 {\phi^\prime}^{\,2}$. This
class of solution is available only if $\phi\neq $ const, more
precisely, if ${\cal W}^\prime+4{\cal W}\ne 0$. A solution
satisfying ${\cal W}^\prime+ 4{\cal W}< 2(y_0-2\sigma_0)$, or with
$\beta>4$, is inflating for all times (cf
figure~\ref{w-vs-eta-latetime}). Let us also restrict the solution
by making the {ans\"atz} $\kappa\phi^\prime\equiv
\sqrt{2/\zeta}\,\e^{-\beta\Delta N/2}$, where $\beta$ is
arbitrary. In this case, the EoS parameter $w$ is less than $-1$
when
\begin{equation}
0< \kappa^2{\phi^\prime}^{\,2} <
\frac{2}{\zeta}\,\frac{y_0-5\sigma_0-3}{5-\beta}.
\end{equation}
If $\phi$ is rolling with a constant velocity, namely $\kappa
\phi^\prime\equiv
 \sqrt{2\phi_0/\zeta}$, then
 $a\propto \e^{\sqrt{\zeta/2\phi_0}\,
(\phi/m_{\rm Pl})}$; the scale factor grows exponentially with
$\phi$. For the power-law expansion, $a(t)\propto t^p$, we get
$\kappa \phi= \sqrt{\frac{2 \phi_0}{\zeta}} \,p \ln t$ and $\kappa
\sigma= \sqrt{\frac{2 \sigma_0}{\gamma}} \,p \ln t$.

\subsection{Absence of potential, $V(\sigma)=0$}

In this subsection, we study the case where there is no potential
for the field $\sigma$, and thus ${\cal Y}=0$. This corresponds to
the case examined previously by Antoniadis, Rizos and
Tamvakis~\cite{Antoniadis93a} (see also
~\cite{Easther:1996a,Kanti:1998jd}), who studied the system
(\ref{GB1})-(\ref{GB4}) numerically.

First, as a special case, we shall consider the solution
$\varepsilon=\varepsilon_0$, or equivalently $a(t)\propto (c_0
t+t_1)^{p}$. The set of three independent equations of motion may
be given by
\begin{eqnarray}
 &\frac{\gamma}{2}\,\kappa^2 {\sigma^\prime}^{\,2} =
\e^{-(5+\varepsilon_0) N}\left[c_1 -\int \e^{(5+\varepsilon_0)
N}\left({\cal W}^\prime+(5+\varepsilon_0)
{\cal W}+3(1+\varepsilon_0)\right) \dd N\right],\nonumber \\
 &\kappa^2 \sigma^\prime \xi_{,\,\sigma} H^2
+\frac{\gamma\kappa^2}{6}\,{\sigma^\prime}^{\,2}= 1 +\frac{{\cal
W}^\prime+(5+\varepsilon_0){\cal W}}{3(1+\varepsilon_0)}, \\
& \kappa^2 \phi^\prime \lambda_{,\,\phi} H^2= \frac{{\cal
W}^\prime+2(3+\varepsilon_0){\cal
W}}{3(1+\varepsilon_0)},\nonumber
\end{eqnarray}
where ${\cal W}\equiv \frac{\zeta}{2}\kappa^2
{\phi^\prime}^{\,2}$, $\varepsilon_0\equiv -1/p$ and $c_1$ is an
integration constant. In particular, for a logarithmic dilaton,
$\phi(t)\propto \ln (c_0t+t_1)$, and thus $\dot{\phi}/H=$ const
$\equiv Q$, we get
\begin{equation}
\frac{\gamma}{2}\,\kappa^2{\sigma^\prime}^{\,2} = c_1\,\e^{(1/p-5)
N}-\frac{\zeta}{2} \kappa^2 Q^2
-\frac{3(p-1)}{5p-1}.\label{no-poten-power-law}
\end{equation}
Let us assume that $c_1$ is positive. There can exist some real
differences between the cases $p>1$ and $p<1$, and also between
$p>0$ and $p<0$. For $1/5<p<1$, the modulus field $\sigma$
normally behaves as a canonical scalar, even if $\zeta>0$, since
$\kappa Q < 1$. While, for $p>1$, and also for $p<0$, since the
last term in (\ref{no-poten-power-law}) is negative, $\sigma$ may
behave as a phantom field, especially for $N\gtrsim 0$.

In the following discussion, we analyze a simple and physically
motivated case where the number of e-folds is proportional to a
shift in field $\sigma$, namely $\kappa \sigma \equiv
\sqrt{2\sigma_0/\gamma}\,N+ $ const \footnote{In the $V(\sigma)=0$
case, the field equations are symmetric in $\phi \to \sigma$,
except that $\xi_{,\,\sigma} \to -\lambda_{,\,\phi}$ (or ${\cal U}
\to - {\cal V}$) and $\gamma \to \zeta$.}.
In this case, the set
of three independent equations of motion reads as
\begin{eqnarray}\label{sol-no-potential}
&(1+\varepsilon) (2\sigma_0+{\cal U})
+4\sigma_0=0, \nonumber \\
& {\cal W}^\prime+4{\cal W}- \frac{4\sigma_0 (2{\cal W}-3{\cal
V})}{2\sigma_0+3{\cal U}}=0,\\
& 3({\cal V}-{\cal U})-{\cal W}-\sigma_0+3=0,\nonumber
\end{eqnarray} where ${\cal U}\equiv \kappa^2 \sigma^\prime H^2
\xi_{,\,\sigma}$ and ${\cal V}\equiv \kappa^2 \phi^\prime
\lambda_{,\,\phi} H^2$. It is not difficult to see that ${\cal
U}=0$ implies only a decelerating universe; the combined effect of
the (scalar) fields is such that the EoS parameter $w\equiv
-1-2\varepsilon/3=1$, that is, the EoS for a stiff matter.
However, with ${\cal U}\neq 0$, there exist other possibilities;
for example, for $|{\cal U}|> 2\sigma_0$, the universe is
accelerating if ${\cal U} <0$. This acceleration can be
super-luminal ($w<-1$), for $0>{\cal U}> -4\sigma_0$; this is
consistent with the numerical results discussed
in~\cite{Antoniadis93a}. Below we shall study two more special
cases.

\begin{figure}
\bc
\includegraphics[width=3.2in]{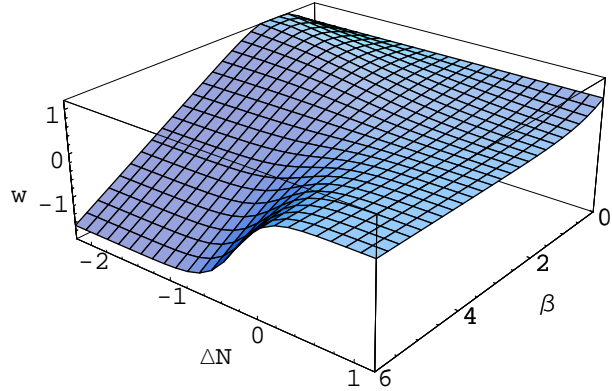}
\ec \caption{\label{w-vs-eta-zero-poten} A non-constant dilaton
case (cf equation~(\ref{sol-no-potential})) with
$\kappa\phi^\prime\propto \e^{-\beta(N-N_0)/2}$. For $\beta>5$,
there exists a crossing between the $w<-1$ and $w>-1$ phases, when
$ N\sim N_0$. The universe is not accelerating for any value of
$\beta$ when $ N\gtrsim N_0$. Here we have chosen $\sigma_0=1$,
but the behaviour of the solution is not changed much in a wide
range of $\sigma_0$ ($>0$). }
\end{figure}

(i) Let us make the {ans\"atz} $\kappa (\dot{\phi}/H) \propto
a^{\beta/2}$, more specifically, $\kappa\phi^\prime\equiv
\sqrt{2/\zeta}\,\e^{-\beta(N-N_0)/2}$, where $\beta$ is an
arbitrary constant. A solution of this type, with $\beta\ge 5$,
may be used for explaining a late-time acceleration of the
universe. More specifically, for $\beta=5$, we find $w\sim -1$
when $ N\lesssim N_0$ but $w>-1/3$ when $ N> N_0$. For $\beta>5$,
there exists a narrow range of $N$ where the EoS parameter $w$
falls between $-1$ and $-1/3$ (cf
figure~\ref{w-vs-eta-zero-poten}).

(ii) Next, let us consider the asymptotic solution $\kappa
\phi(t)\equiv \sqrt{\frac{2\phi_0}{\zeta}}\, \ln (a(t))+$ const.
Then we find
\begin{eqnarray}
& \varepsilon =
-1-\frac{4(\sigma_0+\phi_0)}{3+\sigma_0+\phi_0},\qquad  \kappa H^2
\xi_{,\,\sigma} = \sqrt{\frac{\gamma\sigma_0}{2}}\,
\left(1-\frac{\sigma_0}{3}-\frac{\phi_0}{\sigma_0+\phi_0}\right),\nonumber
\\ \\
& \kappa H^2 \lambda_{,\,\phi} = \sqrt{\frac{\zeta\phi_0}{2}}\,
\frac{\phi_0+\sigma_0-3}{3(\phi_0+\sigma_0)}.\nonumber
\end{eqnarray}
Acceleration is possible only if $-3< \sigma_0+ \phi_0<0$, which
requires that either $\phi_0<0$ or $\sigma_0<0$ (or both). This is
essentially an example of phantom cosmology.

\subsection{Small kinetic terms}

Consider the case where the kinetic term for $\sigma$ is absent or
it is negligibly small, $\dot{\sigma}/H \simeq 0$~\footnote{This
then represents the case of an almost stabilized volume modulus;
the product term $\sigma^\prime \frac{d\xi}{d\sigma}$ is not
essentially vanishing.}. Then there exist two classes of
solutions: the first one is given by
\begin{eqnarray}
&  \phi^\prime
\lambda_{,\,\phi}-\sigma^\prime \xi_{,\,\sigma} =\frac{1}{\kappa^2
H_0^2} \left[\frac{c_1}{3}\,\e^{-4N}+ \frac{c_2}{3}\,
\e^{-2 N}- \e^{2N}\right], \nonumber \\
& V(\sigma) = c_1 \,\frac{H^2}{\kappa^2}\,\e^{2 N }, \qquad
\kappa\phi^\prime = \sqrt{\frac{2c_2}{\zeta}}\,\e^{-2N}.
\end{eqnarray}
This is the critical solution with zero acceleration, $a\propto
t$, or $H=H_0\,\e^{-N}$, implying that $V(\sigma)=$ const. The
second solution may be obtained by solving the equations:
\begin{eqnarray}\label{branch2-no-Kine}
& {\cal W} = \e^{-\int (\varepsilon+5) {\dd  N}} \left[c_3- \int
\e^{(\varepsilon+5)\dd N} \left({\cal
Y}^\prime+(\varepsilon-1){\cal Y}+3(1+\varepsilon)\right)
 {\dd  N}\right],\nonumber \\ \\
& {\cal V}=-1+\frac{{\cal W}}{3}-\frac{{\cal
Y}^\prime+(\varepsilon-1){\cal Y}}{3(1+\varepsilon)}, \qquad {\cal
U}=-\frac{{\cal Y}^\prime+2{\cal
Y}\varepsilon}{3(1+\varepsilon)}.\nonumber
\end{eqnarray}
In the case the kinetic term $\phi$ is also vanishingly small,
$\phi^\prime\simeq 0$, so that ${\cal W}\simeq 0$,
(\ref{branch2-no-Kine}) yields
\begin{eqnarray}
& \fl V(\sigma)=m_{Pl}^2 H^2 \exp\left[{\int(1-\varepsilon)\dd
N}\right]\left(c_3 -3\int (1+\varepsilon)\exp\left[-\int
(1-\varepsilon)\dd N\right] \dd N\right),\nonumber \\ \\
& \fl \sigma^\prime \xi_{,\,\sigma}=
\frac{m_{Pl}^2}{H^2}-\frac{V(\sigma)}{3 H^4}, \qquad \phi^\prime
\lambda_{,\,\phi}=0.\nonumber
\end{eqnarray}
A {prior} knowledge about the time-dependence of the scale factor
helps for retrieving the functional form of the potential
$V(\sigma)$ (or vice versa). For the power-law expansion $H\propto
\e^{\varepsilon_0 N}$, we find
\begin{equation}
V(\sigma) \equiv V_0\,\e^{2\varepsilon_0 N} +
V_1\,e^{(1+\varepsilon_0) N},\quad \xi(\sigma)\equiv \xi_0\,
\e^{-\,2\varepsilon_0 N}+ \xi_1\,\e^{(1-3\varepsilon_0)N},
\end{equation}
where $V_0$ and $\xi_0$ may be (uniquely) determined, but $V_1$
and $\xi_1$ are arbitrary. Alternatively,
\begin{equation}\label{poten-two-parts}
V(\sigma)\equiv V_0\, {a^{-2/\alpha}}+
V_1\,{a^{-(1/\alpha-1)/\alpha}},
\end{equation}
where $\alpha\equiv -1/\varepsilon_0$. This reveals an interesting
possibility that during the matter domination era, $a\propto
t^{2/3}$, the scalar potential may decrease (slowly) as $V\propto
a^{-1/2}$ rather than the usual result $V\propto a^{-3}$.
Moreover, in the presence of matter fields, this behaviour can
change; in particular, with $c_1=0$, one finds $V\propto
a^{-3(1+\bar{w})}$, where $\bar{w}$ characterizes the EoS
parameter of a barotropic fluid.

\subsection{Absence of modulus-GB coupling}

Let us consider that ${\cal U}=0$ (or $\xi(\sigma)=0$) and also
assume that $\sigma$ is rolling negligibly slow,
$\dot{\sigma}\simeq 0$. In the absence of scalar potential,
$V(\sigma)=0$, we find
\begin{eqnarray}
& \fl \zeta{\phi^{\prime}}^{\,2}=\frac{2}{\kappa^2}
\exp\left[-\int (5+\varepsilon)\dd N\right] \left(c_1-
3\int(1+\varepsilon) \exp\left[\int(5+\varepsilon)\dd
N\right]\,\dd N\right), \nonumber \\ \\
&\fl \zeta{\phi^\prime}^{\,2}-6\phi^\prime H^2
\lambda_{,\,\phi}=\frac{6}{\kappa^2},\nonumber
\end{eqnarray}
where $c_1$ is an integration constant. These are the same set of
equations as in (\ref{no-sigma-terms}), so we would not repeat our
analysis for this case. With a non-trivial potential,
$V(\sigma)\ne 0$, the system of equations reduces to
\begin{eqnarray}\label{stab-nonzeroV}
&\zeta{\phi^\prime}^{\,2}=\frac{2}{\kappa^2}\, \sqrt{{\cal Y}}\,
\e^{-5 N}\left[c_1+\int \frac{({\cal Y}-3)(2{\cal Y}-{\cal
Y}^\prime)}{2 {\cal Y}^{3/2}}\,\e^{5 N}\,\dd N\right],\nonumber \\
\\
& \zeta{\phi^\prime}^{\,2}-6\phi^\prime H^2
\lambda_{,\,\phi}=\frac{6}{\kappa^2}-2\Lambda_0,\nonumber
\end{eqnarray}
where ${\cal Y}=\Lambda_0 \exp[-2\int \varepsilon dN]$. Typically,
the rolling of $\sigma$ is proportional to the time variation of
the scalar potential, but with $\dot{\sigma}= 0$, $V(\sigma)=
\Lambda_0/\kappa^2$. Especially, for the power-law expansion,
$H\equiv H_0\,\e^{\varepsilon_0 N}$ ($\varepsilon_0\ne -1$), we
find
\begin{equation}\label{sol-stab}
\frac{\zeta\kappa^2{\phi^\prime}^{\,2}}{2}= c_1
\e^{-(5+\varepsilon_0)
N}-\frac{3(1+\varepsilon_0)}{5+\varepsilon_0}+ c_2\,
\e^{-2\varepsilon_0 N},
\end{equation}
where $c_2$ is another integration constant. One may consider the
possibility that $\varepsilon_0>0$, or $a(t)\propto (t_\infty- c_1
t)^{\alpha}$ with $\alpha<0$, which implies a phantom like
``equation of state", $w<-1$. In the case the numerical
coefficients $c_1$ and $c_2$ are positive, it is possible to
attain $w<-1$ even with a canonical dilaton ($\zeta>0$).

\subsection{Absence of GB terms}

By ignoring the GB couplings, so that ${\cal U}=0$ and ${\cal
V}=0$, one obtains perhaps the most canonical cosmological model
with two scalars. If only $\sigma$ is evolving and $\phi$ is
constant, then of course we find the standard result: ${\cal
Y}=3+\varepsilon$, ${\cal X}=-\varepsilon$. However, if $\phi$ is
also evolving, then these relations are modified as
\begin{equation}
{\cal Y} = 3+\varepsilon, \quad {\cal X}= -\varepsilon -{\cal W},
\quad \varepsilon = -3 -\frac{1}{2}\frac{{\cal W}^\prime}{{\cal
W}}=-3-\frac{\phi^{\prime\prime}}{\phi^\prime}. \label{branch2}
\end{equation}
The implication of the additional field $\phi$ is clear: while the
(functional) form of the potential is not changed, the kinetic
term is modified. Let us discuss the result (\ref{branch2}), in
some detail. At the onset of inflation, $\varepsilon=-1$, one has
$\phi \propto {1}/{a^2}$. During inflation, in at least some
region of field space, $\varepsilon \to 0$ and hence $\phi\propto
{1}/{a^3}$. Clearly, a linear dilaton background, $\phi(t)=\alpha
+ \beta\, t$, which implies $\phi^\prime=\beta/H$ and
$\phi^{\prime\prime}=-\beta H^\prime/ H^2$, is not a solution to
the system of equations (\ref{branch2}), while, a logarithmic
dilaton background, $\phi\propto \ln t$, is clearly a solution.
But this last {\it ans\"atz} implies, in conjunction with
(\ref{branch2}), that $a(t)\propto t^{1/3}$, which is clearly a
non-accelerating solution.

If the system of equations (\ref{branch2}) is to support a
transition between $w< -1$ and $w>-1$ phases, then one needs to
find a solution for $\phi$ such that its evolution takes the ratio
$\phi^{\prime\prime}/\phi^\prime$ from $<-3$ to
$>-3$~\footnote{The requirement of
$\phi^{\prime\prime}/\phi^\prime=
H^{-1}(\ddot{\phi}/\dot{\phi})-\varepsilon <-2$ for a cosmic
acceleration is not to be confused with the slow roll type
condition $|\ddot{\phi}|/\dot{\phi}^2\ll 1$ often used in the
literature.}. The solution $\phi\propto 1/a^\lambda$ with
$\lambda>2$ inflates forever. In particular, for the power-law
expansion, $a\propto (c_0 t+ t_1)^\alpha$ or $H\propto
\e^{-N/\alpha}$, we get
\begin{equation}
\phi = \phi_0 + \phi_1\,\e^{(-3+1/\alpha)
 N},\qquad {\sigma^\prime}^{\,2}= \frac{2}{\alpha\gamma
\kappa^2}-\frac{\zeta}{\gamma}\, {\phi^\prime}^{\,2}.
\end{equation}
It is possible (but not restricted) that $\phi$ and $\sigma$ both
vary as $\kappa \phi\propto \sqrt{\frac{\alpha}{\zeta}}\, \ln t$,
$\kappa \sigma\propto \sqrt{\frac{\alpha}{\gamma}}\, \ln t$; their
combined effect may simply be that $V(\sigma)\propto 1/t^2\propto
a^{-2/\alpha}$. In a matter dominated universe, this yields
$V\propto 1/a^3$ ~\cite{Liddle:1998jc,Ish05-twist}. Of course, in
the absence of GB couplings, there is no effect like
$V(\sigma)\propto 1/\sqrt{a}$ coming from the second term in
(\ref{poten-two-parts}).

To summarize, we have seen that, with the standard approximation:
$H(\sigma)\propto V^{1/2}(\sigma)$, $K(\sigma)\propto H^2(\sigma)$
and $\phi=$ const, the universe accelerates for $V>2K$ with the
equation of state parameter $w>-1$. This result is independent of
the strength of the modulus-GB coupling. If $\phi$ is
non-constant, such as $\phi^\prime \propto \e^{-\beta\Delta N/2}$,
a smooth crossing between $w<-1$ and $w>-1$ phases is possible for
$\beta>4$. If the scalar potential, $V(\sigma)$, is vanishing,
then such a crossing is possible only if $\phi$ is behaving as a
phantom field and/or the quantity ${\cal U}$ ($\equiv\kappa^2
\sigma^\prime H^2 \xi_{, \sigma}$) changes its sign. Another
interesting, as well as phenomenologically safe, example is
$\dot{\sigma}\simeq 0$, that is the case with an almost constant
modulus field. It is possible to attain $w<-1$ in this case by
suitably choosing the integration constants (cf (\ref{sol-stab})),
even if $\phi$ is canonical ($\zeta>0$). In the case when there
are no GB-type corrections, the solutions with a non-constant
dilaton are particularly interesting. The solution
$\phi=\phi_\infty+ \phi_1\,a^{-\lambda}$, with $\lambda>2$,
inflates forever; since
$w=1+\frac{2}{3}\frac{\phi^{\prime\prime}}{\phi^\prime}$,
$\lambda=3$ corresponds to the case of a pure cosmological
constant term, $w_\Lambda=-1$.

\section{Second-order system}

Let us introduce the following set of variables:
\begin{eqnarray}\label{variables2}
&& {\cal X} = \frac{\gamma\kappa^2}{2}
\left(\frac{\dot\sigma}{H}\right)^2, \quad {\cal Y}= \kappa^2
\frac{V(\sigma)}{H^2}, \quad {\cal W} \equiv
\frac{\zeta\kappa^2}{2}
\left(\frac{\dot\phi}{H}\right)^2,\nonumber \\ \\
&&\bar{\mathcal U}=\kappa^2 \xi(\sigma) H^2, \quad \bar{\mathcal
V} =\kappa^2 \lambda(\phi)H^2, \quad  \varepsilon=
\frac{\dot{H}}{H^2}.\nonumber
\end{eqnarray}
Note that $\bar{\mathcal U}$ and $\bar{\cal{V}}$ defined above are
different from those in equation~(\ref{variables1}); here they do
not involve derivatives in their definitions (see the appendix for
additional details). These provide an alternative way to simplify
the study of the model further. Equations (\ref{GB1})-(\ref{GB4})
now form a different set of differential equations:
\begin{eqnarray}
& \fl 0= 3 (\bar{\mathcal U}^\prime-\bar{\mathcal
V}^\prime)-6\varepsilon(\bar{\mathcal U} -\bar{\mathcal V})+{\cal
X}+{\cal W}+ {\cal Y}-3,\label{main1a} \\
&\fl 0=\bar{\mathcal V} ^{\prime\prime}-\bar{\mathcal
U}^{\prime\prime}+(2-\varepsilon) (\bar{\mathcal V}
^\prime-\bar{\mathcal U}^\prime)-2(\bar{\mathcal V} -\bar{\mathcal
U})(\varepsilon^\prime+\varepsilon^2+2\varepsilon)+{\cal X}+{\cal
W}  -{\cal Y} +2\varepsilon+3,
\label{main2a}\\
&\fl 0={\cal X}^\prime + {\cal Y}^\prime+6{\cal X}
+2\varepsilon({\cal X}+{\cal Y}) + 3 (1+\varepsilon)(\bar{\mathcal
U}^\prime-2\varepsilon \bar{\mathcal U}),\label{main3a}\\
 &\fl 0=
{\cal W}  ^\prime + 2(\varepsilon+3) {\cal W} - 3
(\varepsilon+1)(\bar{\mathcal V} ^\prime-2\varepsilon
\bar{\mathcal V} ). \label{main4a}
\end{eqnarray}
Only equation (\ref{main2a}) is second order in derivative, which
can however be discarded since any solution following from the
other three equations will automatically be satisfied by this
equation.


In a constant dilaton background, for which ${\cal W}=0$, the
above system of equations simplify further. There are broadly
three different classes of solutions:
\begin{eqnarray}
(i)&& \quad \bar{\mathcal V} =0, \qquad
\quad \varepsilon=\varepsilon( N),\label{first-branch} \\
(ii) && \quad \bar{\mathcal V} =\bar{\mathcal V} ( N), \qquad
\varepsilon=\frac{\bar{\mathcal V} ^\prime}{2\bar{\mathcal V}
},\label{second-branch} \\
(iii)&& \quad \bar{\mathcal V} =\bar{\mathcal V} ( N), \qquad
\varepsilon=-1.\label{third-branch}
\end{eqnarray}
The first branch, (\ref{first-branch}), corresponding to the case
with no dilaton-dependent terms was analyzed before
in~\cite{Ish:20005sd}. The second branch, (\ref{second-branch}),
implies $ \bar{\mathcal V} \propto H^2$ and thus $\lambda(\phi)=$
const.  The first two branches satisfy
\begin{eqnarray}
&&\frac{\bar{\mathcal
U}^{\prime\prime}}{2}+\frac{5-\varepsilon}{2}\,\bar{\mathcal
U}^\prime- (\varepsilon^\prime +
\varepsilon^2+5\varepsilon)\bar{\mathcal U}=3+\varepsilon-{\cal Y},\\
&&\frac{\bar{\mathcal
U}^{\prime\prime}}{2}-\frac{1+\varepsilon}{2}\,\bar{\mathcal
U}^\prime
 -(\varepsilon^\prime +
\varepsilon^2-\varepsilon)\bar{\mathcal U}=\varepsilon+{\cal X},
\end{eqnarray}
while, the third branch satisfies
\begin{eqnarray}
& \frac{1}{2}(\bar{\mathcal V} ^{\prime\prime}-\bar{\mathcal
U}^{\prime\prime})+3(\bar{\mathcal V} ^\prime-\bar{\mathcal
U}^\prime) +4(\bar{\mathcal V} -\bar{\mathcal U})={\cal Y}-2, \nonumber\\
& \frac{1}{2}(\bar{\mathcal V} ^{\prime\prime}-\bar{\mathcal
U}^{\prime\prime})+2(\bar{\mathcal V} -\bar{\mathcal U})=-{\cal
X}+1.
\end{eqnarray}
This is actually the critical case with zero acceleration.

\subsection{Inflating without a scalar potential}

\begin{figure}
\bc
\includegraphics[width=3.2in]{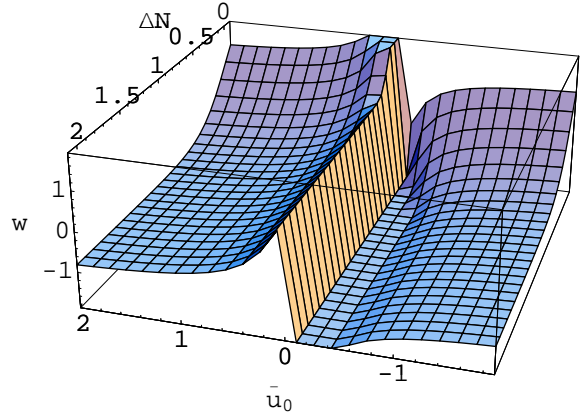}
\ec \caption{\label{figure3new} The EoS parameter $w$ ($\equiv
-1-2\varepsilon/3$) as a function of $\bar{u}\Z0$ and $N$. For
$\bar{u}\Z0 < 0$, we get a super-inflationary solution,
$\dot{H}/H^2>0$, which corresponds to the $\delta<0$ case first
studied in~\cite{Antoniadis93a}. Interestingly, for $\bar{u}\simeq
$ const, it is possible to get a cosmic inflation of an arbitrary
magnitude even for $\delta>0$ (or $ \bar{u}
>0$) without violating the null energy condition. Of course, $w >
-1$ in this case. }
\end{figure}

In this subsection we show that in our model it is possible to
achieve a cosmic inflation without a scalar potential. For
simplicity, let us also drop all $\phi$-dependent terms and work
momentarily in terms of the original variables ($x$, $\bar{u}$ and
$\varepsilon$) and in the units $\kappa=1$. Recall that
\begin{equation}
\bar{u} \equiv \xi(\sigma) H^2
\end{equation}
This quantity should normally decrease with the expansion of the
universe, so that all higher order corrections to Einstein's
theory are only sub-leading~\footnote{Especially, in the case
$\bar{u} \simeq $ const, the coupled Gauss-Bonnet term
$\frac{1}{8}\, \xi(\sigma) {\cal R}_{GB}^2 \equiv 3 \xi(\sigma)
H^4 (1+\varepsilon)= 3 \bar{u}\Z0 H^2(1+\varepsilon)$ may be
varying being proportional to the Einstein-Hilbert term,
$R/2=3H^2(2+\varepsilon)$. One may need to satisfy the condition
$|\bar{u}\Z0| < 1$ at late-times, so that the coupled Gauss-Bonnet
term is not dominating the dynamics; a phenomenologically safe
choice is $|\bar{u}|\lesssim 0.2$. However, there is no such a
restriction for early inflation.}. With $V(\sigma)=0$, the
equations of motion reduce to the form
\begin{equation}\label{Antoni.eta.}
\gamma x^2 = 6 - 6 \bar{u}^\prime + 12 \varepsilon \bar{u}, \quad
\varepsilon^\prime+\varepsilon(\varepsilon+5)=\frac{\bar{u}^{\prime\prime}
+(5-\varepsilon) \bar{u}^\prime}{2 \bar{u}}-
\frac{\bar{u}}{\varepsilon}.
\end{equation}
Fixing the (functional) form of $\bar{u}$ would alone give a
desired evolution for $\varepsilon$. In particular, for $\bar{u}
\simeq $ const $= \bar{u}_0\ne 0$, the explicit solution for
$\varepsilon$ is given by
\begin{equation}\label{soln-varepsilon}
\varepsilon=-\beta_0 + \beta \tanh\left(\beta \Delta N\right),
\end{equation}
where $\Delta N= N+C$, with $C$ being an integration constant, and
\begin{equation}
\beta_0 \equiv \frac{5 \bar{u}\Z0 +1}{2 \bar{u}\Z0}, \quad \beta
\equiv \frac{1}{2 \bar{u}\Z0} \sqrt{25 \bar{u}\Z0^2-2 \bar{u}\Z0 +
1}.
\end{equation} The evolution of $\sigma$ is given by
$\sigma^\prime =\pm \sqrt{(6+12 \varepsilon \bar{u}\Z0)/\gamma}$.
The Hubble parameter is given by
\begin{equation}
H=H_0\,\e^{-\beta_0 N} \cosh(\beta \Delta N).
\end{equation}
For $\bar{u}\Z0 >0$, both $\beta\Z0$ and $\beta$ are positive and
$\beta\Z0
> \beta$, implying that $H$ decreases with the number of e-folds,
and, in turn $w >-1$. However, for $\bar{u}\Z0<0$, we get
$\beta<0$, whereas $\beta\Z0 <0$ for $-0.2< \bar{u}\Z0 <0$ and
$\beta\Z0
> 0$ for $\bar{u}\Z0 <-0.2$. Clearly, with $\bar{u}\Z0 <0$, a solution of
type (\ref{soln-varepsilon}) allows the value $\varepsilon>0$ or
$w<-1$ (cf figure~\ref{figure3new}).

\subsection{Quintessence/phantom crossing}

Within the best-fit concordance cosmology the present
observational data seem to slightly favour a dynamical dark energy
with $w_{de}$ crossing $-1$ from above to below, when the
red-shift factor $z\lesssim 0.6$~\cite{phan-cross-obs}. Could it
be that our universe crosses the phantom divide \footnote{In the
presence of matter fields, $\Omega_m> 0$, one actually defines an
effective EoS for dark energy, $w_{eff}$, whose value can be
(somewhat) different from that of the dark energy EoS parameter,
$w\equiv p_{de}/\rho_{de}$. Nevertheless, if a (fundamental)
theory is to support the crossing of phantom divide, $w=-1$, then
a similar effect is expected in the absence of matter since a
gravitationally attractive source can only make the dark energy
source less squishy or the EoS parameter more positive, $w_{eff}
> w_{de}$ .}, $w=-1$, only recently?
And could it be due to a GB-type modification of general
relativity? This is indeed an interesting subject of recent
discussion on both fronts: observational~\cite{phan-cross-obs} and
theoretical ~\cite{phan-cross-th} cosmology. Though a more
detailed analysis of the model, accommodating the effects due to
matter fields, is needed yet, first indications are that such a
possibility cannot be denied.

First we consider the case where the modulus field $\sigma$ is
behaving similarly to the dilaton $\phi$, so $\bar{\mathcal
V}=\bar{\mathcal U}$. Then there are again three different classes
of solutions:
\begin{eqnarray}
(i) \quad & {\cal X}=-\varepsilon, ~~ \qquad {\cal
Y}=3+\varepsilon,~~ \qquad \bar{\mathcal U}=0,\label{new-branch1} \\
(ii) \quad & {\cal X}=-\frac{\bar{\mathcal
U}^\prime}{\bar{\mathcal U}}, \qquad {\cal
Y}=3+\frac{\bar{\mathcal U}^\prime}{2\bar{\mathcal U}},\qquad
\varepsilon=\frac{\bar{\mathcal U}^\prime}{2 \bar{\mathcal U}},\label{new-branch2}\\
(iii) \quad &  {\cal X}=1, \qquad {\cal Y}=2, \qquad
\varepsilon=-1.\label{new-branch3}
\end{eqnarray}
The first branch is the standard one, implying that
$\lambda(\phi)=0=\xi(\sigma)$, while the third branch gives zero
acceleration, so it is the second branch which is most
interesting. For $H\propto \left(\xi(\sigma)\right)^{-1/2}$, so
that $\bar{\mathcal U}=$ const, the effect is similar to that of a
pure cosmological constant term, $\varepsilon = 0$. Instead, if
$H\propto \left(\xi(\sigma)\right)^{-1/2} a^{-\,\alpha/2}$, then
the universe is accelerating for $\alpha<2$, whereas $\alpha=2$
marks the boarder between acceleration and deceleration.
Typically, for the solution $\bar{\mathcal U}= c_1 N+c_2$ (with
$c_i>0$) we find $\varepsilon <0$ for $N< - c_2/c_1$ and
$\varepsilon
>0$ for $N> - c_2/c_1$, where $N\equiv \ln (a(t)/a_0)$, with $a_0$
being the present value of $a(t)$, and $1+z\equiv \e^{\Delta N}$.
For instance, if $c_2/c_1=0.47$, then the EoS parameter $w$ is
less than $-1$ for $z<0.6$.

\subsection{Deceleration/acceleration phase and crossing of phantom divide}

\begin{figure}
\bc
\includegraphics[width=3.2in]{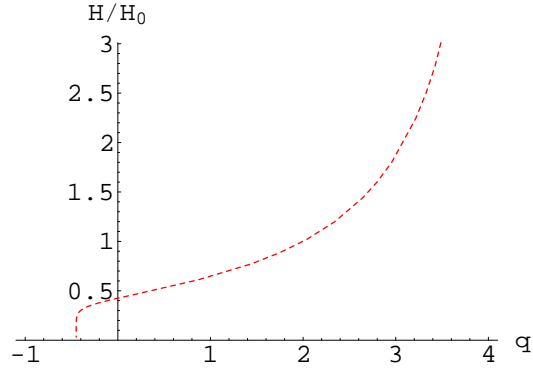}
\ec \caption{\label{no-coupling-first} The universe enters into an
accelerating phase ($\varepsilon>-1$) when $\Delta N \simeq 0.22$,
or $H/H_0\simeq 0.42$ (cf equation~(\ref{branch1-noGB})). For
$\Delta N\gtrsim 1$, the field $\phi$ is almost constant, with the
deceleration parameter $q\equiv -1-\varepsilon \simeq -0.45$ and
$w\simeq -0.63$. }
\end{figure}

Consider the possibility that the coupling functions $\xi(\sigma)$
and $\lambda(\phi)$ both vary inversely with $H^2$, so that
$\bar{\mathcal U}\equiv u_0$ and $\bar{\mathcal V} \equiv v_0$.
For $u_0=v_0$ (which again implies a trivial GB coupling), there
are three different classes of solutions. The first class of
solution is given by
\begin{eqnarray}\label{branch1-noGB}
& \varepsilon = -3+\sqrt{6}\tanh \sqrt{6}\, \Delta N, \quad V
\propto \e^{-\,6N}\,\sinh 2\sqrt{6}\,
\Delta N,\nonumber\\
& \gamma {\sigma^\prime}^{\,2} =
\frac{2}{\kappa^2}\,\left[(6v_0-1)\tanh \sqrt{6}\,\Delta N
- 3(5 v_0-1)\right], \\
& \zeta {\phi^\prime}^{\,2} = -\frac{6}{\kappa^2}\, v_0
\left(2\sqrt{6}\tanh \sqrt{6}\,\Delta N -5\right),\nonumber
\end{eqnarray}
where $\Delta N\equiv N+ N_1$. This solution yields only a
standard inflation for which $\varepsilon<0$ (cf
figure~\ref{no-coupling-first}). The transition from the
deceleration phase to the acceleration phase (or vice verse) is a
natural ingredient of the above solution. In fact, we would not
have gained such an insight into a functional from of
$\varepsilon$ (and the EoS parameter $w$), have we had dropped the
GB couplings from the action, which corresponds to the case
$v_0=0$.

For $|\Delta N| \gtrsim 1$, since $|\tanh \sqrt{6} \Delta N|
\simeq 1$, both the fields $\sigma$ and $\phi$ are rolling with
almost constant velocities, viz, $\sigma\equiv \sigma_0 \sqrt{6}\,
N$, where $\sigma_0\equiv m_{Pl}\sqrt{(2-9v_0)/3\gamma}$. The
scalar potential and the modulus GB coupling take the following
form
\begin{eqnarray}
V(\sigma)&=& V_0\,\e^{-\,\sqrt{6}(\sigma/\sigma_0)}\, \sinh 2
\left(c+ \sigma/\sigma_0\right),\nonumber\\
 \xi(\sigma) &=&
\xi_0\, \e^{\sqrt{6}(\sigma/\sigma_0)}\,\left[\cosh
\left(c+\sigma/\sigma_0\right)\right]^{-\,2},
\end{eqnarray}
where $c$ is an arbitrary constant. A similar expression may be
given for $\lambda(\phi)$, in terms of $\phi$.

\begin{figure}
\bc
\includegraphics[width=3.2in]{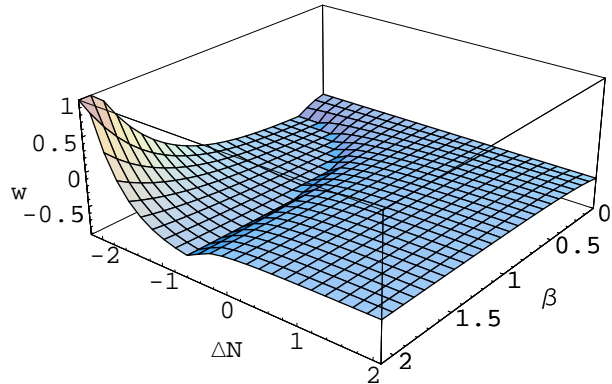}
\ec \caption{\label{zero-coupling-second-a} The equation of state
parameter $w~ (\equiv -1-2\varepsilon/3)$ with
$\kappa\phi^\prime\equiv \sqrt{\frac{2\phi_0}{\gamma}}\,
\e^{-\beta(N- N_0)/2}$ and $\phi_0/v_0=0.2$ (cf  the negative root
in equation~(\ref{branch2-noGB})). Only a small value of $\beta$
seems to be reasonable, otherwise $w$ can be imaginary for $\Delta
N\ll 0$. }
\end{figure}

\begin{figure}
\bc
\includegraphics[width=3.2in]{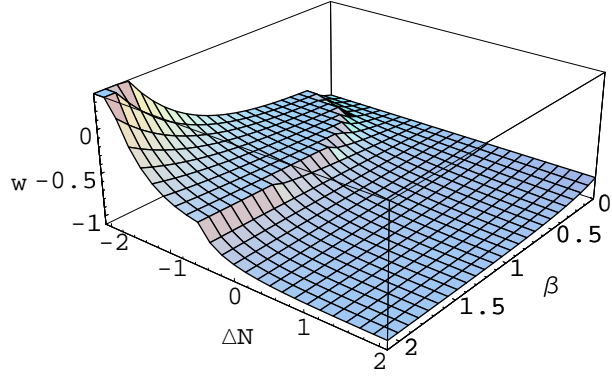}
\ec \caption{\label{zero-coupling-second-b} The equation of state
parameter $w$ as in figure~\ref{zero-coupling-second-b}, but for
the positive root in equation~(\ref{branch2-noGB}). For $\beta>0$,
the universe passes from a decelerating phase to an accelerating
phase, but does not cross the phantom divide, $w=-1$. }
\end{figure}

\begin{figure}
\bc
\includegraphics[width=3.2in]{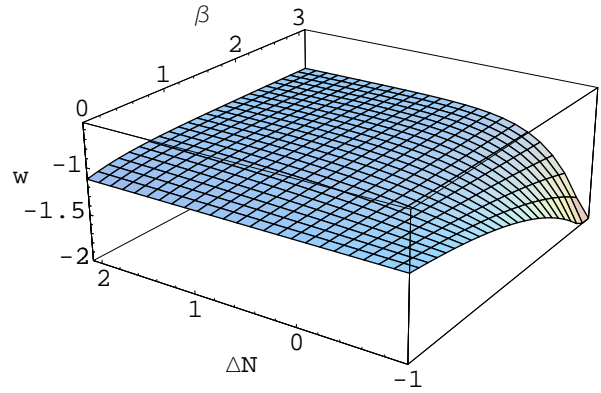}
\ec \caption{\label{zero-coupling-second-c} The equation of state
parameter $w$ with $\kappa\phi^\prime\equiv
\sqrt{\frac{2\phi_0}{\gamma}}\, \e^{-\beta(N- N_0)/2}$ and
$\phi_0/v_0=-0.2<0$ (cf  the negative root in
equation~(\ref{branch2-noGB})). Here $w$ may start from a value
(well) below $-1$ but it never crosses it, leading always to a
phantom phase. }
\end{figure}

\begin{figure}
\bc
\includegraphics[width=3.2in]{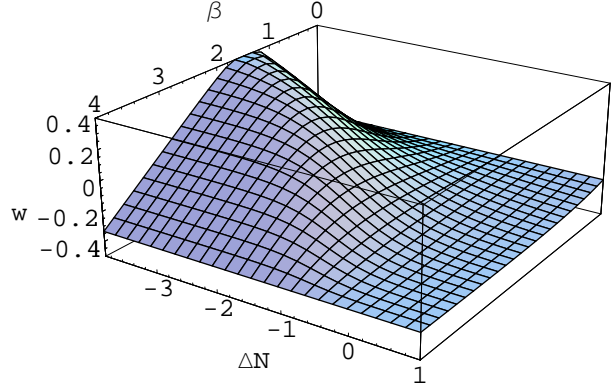}
\ec \caption{\label{zero-coupling-second-d} The equation of state
parameter $w$ as in figure~\ref{zero-coupling-second-c}, but for
the positive root in equation~(\ref{branch2-noGB}). For any value
of $\beta$, the universe is not accelerating, since in this case
$w$ always takes a value larger than $-1/3$. }
\end{figure}

The other two classes of solutions may be described by the
following system of equations:
\begin{equation}
{\cal X}=-\varepsilon-{\cal W}  ,\quad {\cal Y}=3+\varepsilon,
\end{equation}
where, as above, ${\cal W}\equiv \frac{\zeta}{2}\kappa^2
{\phi^\prime}^{\,2}$, and
\begin{equation}\label{branch2-noGB}
\varepsilon = -\frac{1}{2}-\frac{{\cal W} }{6 v_0}\pm \frac{1}{6}
\sqrt{9+\left(\frac{{\cal W} }{v_0}\right)^2 -\frac{6{\cal W}
^\prime}{v_0}-\frac{30{\cal W} }{v_0}}.
\end{equation}
In the constant dilaton case, $\phi^\prime=0$, we find $w=-1$ or
$0$. We assume that we are dealing with a canonical dilaton, so
${\cal W} > 0$. For ${\cal W}/\bar{\mathcal V}_0>0$, the both
roots in (\ref{branch2-noGB}) give the EoS parameter $w>-1$.
However, for ${\cal W}/\bar{\mathcal V}_0<0$, the positive root
can give $w<-1$.

To simplify the analysis further, let us make the {ans\"atz}
$\kappa\phi^\prime\equiv \sqrt{\frac{2\phi_0}{\gamma}}\,
\e^{-\beta\Delta N/2}$, where $\beta$ is arbitrary. Then, for
$\phi_0/v_0>0$, in the regime $ \Delta N \gg 0$, such that
$\phi^\prime/\phi_0\to 0$, the positive root in
(\ref{branch2-noGB}) yields $w\sim -1$ (except for $\beta\simeq
0$), while the negative root yields $w\sim  -1/3$ (see figures
\ref{zero-coupling-second-a} and \ref{zero-coupling-second-b}). On
the other hand, for ${\cal W}/v_0<0$, the positive root supports a
phantom phase ($w<-1$), while the negative root solution is not
accelerating, as it approaches $w=-1/3$ from above (see figures
\ref{zero-coupling-second-c} and \ref{zero-coupling-second-d}). It
is generally for $u_0<0$ (or equivalently $v_0<0$ or $\delta<0$)
that one of the branches can give a non-standard ($w<-1$) or
super-inflationary phase satisfying $p+\rho<0$, in consistent with
the numerical results in ~\cite{Antoniadis93a,Easther:1996a}.

For the power-law expansion, $a(t)\propto t^\alpha$, so that
$H=\alpha/t$, and $\phi^\prime = $ const (or $\phi\propto \ln t +$
const), we require $\phi_0/v_0 <0.30$ or $>29.69$ so as to keep
$\varepsilon$ real. We shall be interested in the case where $v_0$
is small, so that the GB contribution is only sub-leading. For
instance, if ${\cal W} \simeq -0.1 v_0>0$, then $\varepsilon\simeq
-1.06$ or $0.09$; only the second branch is accelerating which
gives $w\simeq -1.06$.

In the $u_0\neq v_0$ case, so that the GB coupling is
non-vanishing, the first branch, (\ref{branch1-noGB}), is modified
as
\begin{eqnarray}
&& \gamma {\sigma^\prime}^{\,2} = \frac{2}{\kappa^2}\left[(7 u_0-
v_0+1)(3+\varepsilon)- 3(6 u_0-v_0-1)\right],
\nonumber \\
&& V = m_{Pl}^2 (v_0-u_0+1)(3+\varepsilon)\,H^2
\end{eqnarray}
without changing the forms of ${\cal W}$ and $\varepsilon$. As for
the second branch, (\ref{branch2-noGB}), the expressions of
$\gamma {\sigma^\prime}^{\,2}$ and $V(\sigma)$ are now more
complicated and not illuminating at least to write down here.
However, a simplification occurs in the case that $H\equiv
H_0\,\e^{\varepsilon_0 N}$ or equivalently, $a\propto (c_0
t+t_1)^\alpha$ ($\alpha>0$) or $a\propto (t_\infty-t)^{-\alpha}$
($\alpha<0$). Especially, the time variations of the scalar fields
are given by
\begin{eqnarray}
&\fl \gamma {\sigma^\prime}^{\,2} = \frac{2}{\kappa^2}
\bigg[-c_1\e^{-2(3+\varepsilon_0) N}
+\frac{3(1+\varepsilon_0)\varepsilon_0 u_0}{3+\varepsilon_0} +
\varepsilon_0(1-\varepsilon_0)(u_0 - v_0)-\varepsilon_0
\bigg],\nonumber\\
&\fl \zeta {\phi^\prime}^{\,2} = \frac{2}{\kappa^2} \left[c_1
\e^{-2(3+\varepsilon_0) N}-\frac{3(1+\varepsilon_0)\varepsilon_0
u_0}{3+\varepsilon_0}\right].
\end{eqnarray}
Here either value of $\varepsilon_0$ is possible; for
$\varepsilon_0>0$, $\phi$ may behave as a phantom field,
especially, if $N>0$ and/or $c_1<0$, unless that $u_0$ (or
$\xi(\sigma))$ is negative.

\subsection{Late-time approximation: two scalar case}

Finally, as a good approximation at late times, consider that one
of the fields, mainly $\sigma$, is rolling very slow such that
$\kappa |\sigma^\prime|\equiv \sqrt{2 x_0/\gamma}$ and $\kappa
H^{-1}(\sigma)\equiv \sqrt{y_0}\, V^{-1/2}(\sigma)$, where $x_0$
is a small constant, while $y_0>0$. In the case of a vanishing
effective GB coupling, $\bar{\mathcal U}=\bar{\mathcal V}$, the
system of equations reduce to
\begin{eqnarray}
&& H^2=\frac{1}{\kappa^2 \xi(\sigma)} \left[c_1\,\e^{-2(3-y_0)
N}-\frac{y_0(x_0+ y_0-3)}{3(y_0-2)(3- y_0)}\right],
\nonumber\\
&& \zeta {\phi^\prime}^{\,2}=\frac{2}{\kappa^2} \left(3-
x_0-y_0\right), \quad \varepsilon=y_0-3.
\end{eqnarray}
The universe accelerates for $y_0>2$. An especially interesting
range for $y_0$ is $2<y_0 <3$, for which the universe naturally
undergoes a cosmic acceleration and the couplings $\xi(\sigma)$
and $\lambda(\phi)$ both decrease exponentially with the number of
e-folds, $ N$.

In the general case, there are five different branches of
solutions: except the first branch, which is giving by
\begin{eqnarray}
&\fl \xi(\sigma)= 0, \quad
\varepsilon=-\frac{3 x_0}{x_0+ y_0},\nonumber\\
&\fl \zeta {\phi^\prime}^{\,2} = \frac{2}{\kappa^2}
\left[c_1\,\e^{-(2+\varepsilon)
N}+\frac{(2 x_0-y_0)(3- x_0-y_0)}{2 x_0+5 y_0}\right], \\
&\fl  H^2 = \frac{1}{\kappa^2 \lambda(\phi)}\,
 \e^{-6 x_0 N/(x_0+ y_0)} \bigg[
c_2-\int \e^{6 x_0 N/(x_0+y_0)}\left(3-x_0-
y_0-\frac{\zeta}{2}\,\kappa^2 {\phi^\prime}^{\,2}\right)\dd
N\bigg],\nonumber
\end{eqnarray}
their expressions are lengthy and not illuminating at least to
write down here. However, especially, for $\varepsilon\equiv
\varepsilon_0$, they merge to give a single solution
\begin{eqnarray}
\zeta{\phi^\prime}^{2}&=&\frac{2}{\kappa^2}\left[
\frac{y_0(1-\varepsilon_0)-3(1+\varepsilon_0)}{5+\varepsilon_0}-x_0
-3 \lambda_1 (5+3\varepsilon_0) \right], \nonumber\\
\xi(\sigma) &=& \frac{1}{\kappa^2 H_0^2}\left( \, \xi_0 +\xi_1
\,\e^{-\,2\varepsilon_0 N} \right),
 \\
\lambda(\phi)&=&\frac{1}{\kappa^2 H_0^2}\left(\, \lambda_0 +
\lambda_1\,\e^{-(5+3\varepsilon_0) N} + \lambda_2\,
\e^{-\,2\varepsilon_0 N}\right),\nonumber
\end{eqnarray}
where $\xi_0$, $\lambda_0$ and $\lambda_1$ are integration
constants, and
\begin{eqnarray}
\xi_1 &\equiv& \frac{3 x_0+(x_0+y_0)\varepsilon_0}{3\varepsilon_0
(1+\varepsilon_0)}, \nonumber \\
\lambda_2 &\equiv&
\frac{(3+\varepsilon_0)[(3+x_0+y_0)\varepsilon_0+3+5
x_0-y_0]}{3\varepsilon_0(5+\varepsilon_0)(1+\varepsilon_0)}
\end{eqnarray}
with $\varepsilon_0\neq -1$. One may replace $N$ by
$\sqrt{\frac{\gamma}{2 x_0}}\,\kappa \sigma$ so as to retrieve a
functional form for $\xi(\sigma)$. Especially, the cases
$\varepsilon_0=0, -5 $ (and also $\varepsilon_0=-5/3$) need to be
treated separately, for which new terms appear with $
N$-dependence. For instance, for $\varepsilon_0=0$ (i.e. $H\equiv
H_0$), we find
\begin{eqnarray}\label{eps=0-special}
&\zeta {\phi^\prime}^{\,2} = \frac{2}{\kappa^2}\left[
\frac{y_0-3}{5} - x_0 + c_1\,
\e^{-\,5 N}\right],\nonumber\\
&\lambda(\phi)= \frac{1}{\kappa^2 H_0^2} \left[\lambda_0
-\frac{\lambda_1}{15}\, \e^{-\,5
N}-\frac{2}{5}(5 x_0+3-y_0) N\right], \\
&\xi(\sigma) = \frac{1}{\kappa^2 H_0^2} \left[\xi_0- 2 x_0
N\right]\equiv \frac{1}{\kappa^2 H_0^2}\, \left[\xi_1-
\sqrt{2\gamma x_0}\,\kappa\sigma\right].\nonumber
\end{eqnarray}
With no surprise, this solution behaves all the times as a
cosmological constant term, for which $w= -1$, since $H\equiv H_0$
and $V(\sigma)=m_{Pl}^2 H_0^2 y_0^2\equiv \Lambda$.

We conclude this section by providing an overview of the different
solutions discussed above. In subsection 5.1 we showed the
existence of an inflationary solution with a vanishing potential
and for the modulus-GB coupling $\xi(\sigma)\propto H^{-2}$. This
result is quite interesting. It is also obvious from the above
analysis that solutions with trivial scalar-GB couplings
($\bar{\cal U}=0=\bar{\cal V}$) are not equivalent to solutions
where these coupling are (negligibly) small, at late times. The
simplest way to visualize this is to set only $\tilde{\cal
U}=\tilde{\cal V}$, i.e., the dilaton $\phi$ is behaving as the
modulus $\sigma$. Unlike in the $\bar{\cal U}=0=\bar{\cal V}$
case, we now find three different kinds of solutions, one of
which, namely (\ref{new-branch2}) can support a crossing between
$w<-1$ and $w>-1$ cosmologies. When $\lambda(\phi)$ and
$\xi(\sigma)$ both grow inversely with the squared of the Hubble
parameter, $H^2$, then, for the first branch, the EoS parameter
$w$ can take a value less than $-1/3$, namely $w\simeq -0.63$ at
late times. Either both of the other two branches support an
inflationary phase with $w>-1$, or only one of the branches
supports an inflationary phase, but with $w<-1$. In the latter
case, the second branch is not accelerating since
$\varepsilon<-1$.

As in the standard picture, with $H \propto V^{1/2}$ and
$\dot{\sigma}\propto H(\sigma)$, the universe accelerates for
$y_0\equiv \kappa^2 \frac{V(\sigma)}{H^2}>2$, even if $\bar{\cal
U}=\bar{\cal V}\ne 0$. This result is independent of the ratio
$\dot{\sigma}/H$. It is the case for $y_0<3$, the function
$\bar{\cal U}$ decreases with the expansion of the universe. In
the general case, for which $\bar{\cal U}\neq \bar{\cal V}\neq 0$,
there are five branches of solutions, which simplify, for
instance, if we take the limit where the slow-roll type variable
$\varepsilon$ is constant.

\section{Runway dilaton and matter coupling}

In a more realistic scenario, $V(\sigma)$ in (\ref{dilatonGB2})
may be replaced by a general form of the potential, namely
$$
V(\sigma)\to V(\phi,\sigma)=\tilde{\lambda}(\phi) V(\sigma), $$
which is now a function of both the modulus $\sigma$ and the
dilaton $\phi$. In our analysis we considered only the simplest
case where $\phi$ was expected to evolve (relatively fast) towards
its finite value $\phi_m$. In this limit,
$\tilde{\lambda}(\phi)\simeq {\rm const}$.

Generally, $\phi$ (the spin-$0$ partner of spin-$2$ graviton) is
coupled to Einstein term and thus, while going from string frame
to Einstein frame, it may also couple to matter fields. In the
heterotic string theory studied by Antoniadis, Gava and
Narain~\cite{Narain92a}, there is no coupling between the field
$\sigma$ and the Ricci-scalar term, so $\sigma$ couples to gravity
only via spacetime curvature tensors or Riemann curvature
invariants. Thus a conformal transformation does not necessarily
involve a $\sigma$-dependent term, but only a dilaton dependent
term. In addition, $\sigma$ generally has a non-zero vacuum
expectation value and hence a non-zero mass~\cite{Antoniadis93a};
it is thus conceivable to expect only a minimal coupling between
$\sigma$ and the matter. Furthermore, in a de Sitter
(accelerating) universe, sypersymmetry has to be broken in
general. If so, dilaton acquires a large enough mass
($m_\phi\gtrsim 10^{-3} {\rm eV}$) after supersymmetry breaking.
Within this scenario, $\phi$ takes an almost constant value and
hence it decouples from matter, so called {\it least coupling
principle}~\cite{Damour:2002a,Damour:1994a}. In any case, let us
assume that $\phi$ is coupled to matter fields non-minimally:
\begin{equation}\label{dilatonGB3}
\fl S = \int d^{4}{x}\sqrt{-g}\bigg[ \frac{1}{2\kappa^2}
R-\frac{\gamma}{2} (\nabla\sigma)^2-V(\sigma) -\frac{\zeta}{2}
(\nabla\phi)^2 + \frac{f}{8}\, {\cal R}^2_{GB} +
A^4(\phi)\left(\rho_M+\rho_R\right) \bigg],
\end{equation}
where $f\equiv [\lambda(\phi)- \delta\xi(\sigma)]$. The equations
of motion that describe gravity, the scalars, and the matter and
radiation are given by (in units $\kappa=1$)
\begin{eqnarray}
&\fl H^2= \frac{1}{3}\left(\frac{\gamma}{2}\dot{\sigma}^2
+\frac{\zeta}{2}\dot{\phi}^2+V
-3 H^3 \dot{f}+A^4 (\rho_M+\rho_R)\right),\label{GB1-new}\\
&\fl \frac{\ddot{a}}{a}=-\frac{1}{3}\bigg[\gamma\dot{\sigma}^2+
\zeta\dot{\phi}^2-V +A^4 \rho_R +\frac{1}{2} A^4 \rho_M
+\frac{3}{2}\, H^2\left(H\dot{f}+\ddot{f}+2
\dot{f}\dot{H}/H\right)\bigg],\\
&\fl
 \gamma(\ddot{\sigma}+3H\dot{\sigma})=
-V_{,\,\sigma}-\frac{1}{8}\,\xi_{,\,\sigma}\,{\cal R}^2_{GB},
\label{GB3-new}\\
&\fl \zeta(\ddot{\phi}+3H\dot{\phi})=-A_{,\,\phi} A^3\,\rho_M +
\frac{1}{8}\,\lambda_{,\,\phi} \,{\cal R}^2_{GB}, \label{GB4-new}
\end{eqnarray}
In the standard scenario of no matter-scalar coupling, e.g. the
model of quintessence, $A(\phi)=1$. For $A(\phi)\neq 1$, the fluid
equation of motion for matter (or radiation) is
\begin{equation}\label{modified-E-cons}
a\frac{\partial\rho_i}{\partial a} + \frac{1}{\tilde{\alpha}}
\,\frac{\partial\rho_i}{\partial\phi} =-3(\rho_i+p_i),
\end{equation}
where $i=M({\rm matter})$ or $R({\rm radiation})$ and
$\tilde{\alpha}\equiv \frac{d \ln A}{d\phi}$. The implicit
assumption is that matter couples to $A^2(\phi) g_{\mu\nu}$ (with
scale factor $\hat{a}$) rather than the Einstein metric
$g_{\mu\nu}$ alone, and $\rho_R\propto 1/\hat{a}^{4}$, where
$\hat{a}\equiv a A(\phi)$ so only $\rho_M$ enters the $\phi$
equation of motion. We can practically drop the $\rho_R$ term, at
late times, since most of the radiation energy in the universe is
in the cosmic microwave background, which makes up a fraction of
roughly $5\times {10}^{-5}$ of the total density of the universe.

In the case $A(\phi)=1$, one can solve the system of equations
(\ref{GB1-new})-(\ref{modified-E-cons}), for example, by making an
ans\"atz for the evolution of one of the scalars and/or the scale
factor of the universe. Note that in this case the energy
conservation equation (\ref{modified-E-cons}) is easily expressed
as a differential equation, namely $\Omega_i^\prime+2\varepsilon
\Omega_i + 3(1+w_i)\Omega_i=0$, where $\Omega_i\equiv
\rho_i/3H^2$, $w_i\equiv p_i/\rho_i$ and $\Omega^\prime=\dd
\Omega/\dd \ln[a(t)]$. In the case $A(\phi)\ne 1$, the extra
(matter) degree of freedom added to the model complicates the
problem, in which case the field equations may be solved
numerically~\cite{Ish06b}.

Especially in the case where $\phi$ evolves considerably slow with
time, $A(\phi)$ is not essentially a constant, even at late times.
In this case, a question may be raised as to whether the
equivalence principle (universality of free fall in a given
gravitational field) is violated due to a runway
dilaton~\cite{Damour:2002a,Damour:1994a}, leading to time
variations of some of the fundamental constants, like the fine
structure constant $\alpha_e=e^2/\hbar c$. For metrically coupled
matter theories, one may also raise a question regarding the time
variation of Newton's constants under post-Newtonian
approximation~\cite{Damour:1990a}. But the constraints on
gravitational actions of the type considered here, under
(post-)Newtonian approximations, may be less stringent as they
only require~\cite{Farese:2004cc,Amendola:2005} (in units
$\kappa=1$)
\begin{equation}\label{post-New}
 \dot{\sigma} H^2 \frac{d\xi(\sigma)}{\dd\sigma} <
{10}^{-2},\qquad \dot{\phi} H^2 \frac{\dd\lambda(\phi)}{\dd\phi} <
{10}^{-2}.
\end{equation}
In our notations, introduced in equation (\ref{variables1}), these
read ${\cal U} H< {10}^{-2}$ and ${\cal V} H < {10}^{-2}$. This
result does not rule out any scalar-GB couplings. Within our
model, both ${\cal U}$ and ${\cal V}$ may be exponentially small
(close to zero). In any case, if ${\cal U}\sim H^2
\frac{d\xi(\sigma)}{\dd\sigma}\sim {\cal O}(1)$ and ${\cal V}\sim
H^2 \frac{d\lambda(\phi)}{\dd\phi}\sim {\cal O}(1)$, then we would
require that $\dot{\sigma}, \dot{\phi}< {10}^{-2}$ (in units
$\kappa=1$).

If $\lambda(\phi)$ and $\xi(\sigma)$ are nearly constants,
especially, after inflation, which we assume to be the case here,
then action (\ref{dilatonGB3}) may be compared to the
gravitational action considered in \cite{Damour:2002a} (cf
equation~(2.4)), under the replacement, $V(\sigma)\to
V(\phi,\sigma) =\tilde{\lambda}(\phi) V(\sigma)$. For most of the
solutions that we have presented in this paper, after a certain
number of e-folds of expansion, the rolling of $\phi$ could be
small as it is given by $\phi^\prime= \sqrt{\frac{2{\cal
W}}{\zeta}} H \sim \e^{- \,\beta N}$ with $\beta N>0$. In a
specific model considered in~\cite{Damour:2002a},
$\tilde{\lambda}(\phi)\equiv \lambda_\infty (1+b_\lambda
\e^{-\,c\phi})$, for $ \frac{\dot{\phi}}{H}=\phi^\prime < 0.7$,
the level of variation of fine structure constant, $|d \ln
e^2/dt|\lesssim 10^{-21} {\rm yr}^{-1}$ is much smaller than the
current best limit on the time variation of $e^2$, namely $|d \ln
e^2/dt|\lesssim 10^{-17} {\rm yr}^{-1}$; thus, a small rolling of
$\phi$ is physically viable and phenomenologically acceptable.

One may be more concerned here with a time-variation of the
(common) modulus $\sigma$. We take $\dot{\sigma}=0$ as a
conservative case, for which $V(\sigma)$ is also constant (cf
equation~(\ref{stab-nonzeroV})). Because of phenomenological
reasons it is desirable that the rolling of $\sigma$ is small,
especially after inflation, which then allows only a slow rolling
of the potential. Instead of the condition $\dot{\sigma}=0$, we
expect $\kappa \dot{\sigma}< H$ to hold at all times, i.e.
$\sqrt{2\cal X}< \sqrt{\gamma}$. This last condition is satisfied
by all our solutions discussed in the previous sections.

In has been known that even if the tests of general relativity in
the solar system (and binary pulsars) set strong constraints on
scalar-tensor models of the type considered in this paper, they
may differ significantly from GR at high red-shift, $z\gtrsim
0.7$~\cite{Farese:2000a,Farese:2004cc}, possibly affecting the
standard scenario of structure formation. In turn, the scalar-GB
couplings, in addition to the matter-dilaton coupling, may be
constrained only by taking into account cosmological and
solar-system data together~\cite{Farese:2004cc}). Such results are
not known yet; see also the recent paper~\cite{Koivisto:2006}
which discusses on cosmological and astrophysical constraints of
Gauss-Bonnet dark energy.

For $A(\phi)\simeq 1$ and $f\simeq 0$, one has $\dot{\phi}\propto
1/a^3$ (cf equation~(\ref{GB4-new})); with the large scale factor,
the rolling of $\phi$ can be small. In fact, the universe has
merely doubled in its size between radiation domination era
($t\sim {10}^{11} ~{\rm sec}$) and today ($t\sim {10}^{17}~{\rm
sec}$) or $N=1+z \sim 2$, where $z$ is the red-shift factor. So,
unless we do not demand a large number of e-folds (between
radiation domination and present eras), we find no reason to
expect a large shift in $\phi$, since $\Delta\phi\equiv
\sqrt{\frac{2}{\zeta}}\int \sqrt{\cal W} H \dd t\simeq
\sqrt{\frac{2{\cal W}_0}{\zeta}}\,\Delta N$. It may well be that
the field $\phi$ has evolved by a small amount; provided
$A_{,\,\phi}/A < 10^{-3}$, or more mildly
$\tilde{\alpha}^2<{10}^{-5}$~\cite{Martin:2005}, for today's value
of $\phi$, any possible violation of equivalence principle (or
deviations from the standard general relativity) may be small
enough to easily satisfy all cosmological constraints.

\medskip
\section{Conclusions}

In this paper, we explored the cosmological implications of the
modulus-dilaton dependent one-loop corrections to the
gravitational coupling of the heterotic string effective action,
by presenting various exact cosmological solutions. These
corrections comprise a GB invariant multiplied by universal
non-trivial function of the common modulus field $\sigma$ and the
dilaton field $\phi$, and thus they may have an important role in
explaining inflation, as well as the recently discovered
acceleration of the cosmic expansion. The solutions represent a
flat four-dimensional FRW spacetime where the Hubble expansion
parameter is a smoothly decreasing function of the number of
e-folds, $N$, while remaining non-zero at all times. As such, they
exhibit the type of branch-changing solution, i.e.,
deceleration/acceleration regime, that is a prerequisite for the
successful incorporation of the early inflation as well as the
currently observed deceleration/acceleration phase in the
expansion of the (late) universe.

The main purpose of this paper was to point out the importance of
$\alpha^\prime$-corrections for explaining the cosmic acceleration
problem attributed to "dark energy" or gravitational vacuum
energy. It may well be that the evolution of a cosmological (FRW)
background from the string perturbative vacuum leads to a regime
of cosmic inflation in which the strengths of the coupling
constants associated with the string loop effects can grow with
time, though very slowly, while the higher curvature contributions
to the effective action are more suppressed than the Ricci scalar.
When both the common modulus filed ($\sigma$) and the dilaton
($\phi$) take almost constant values at late-times, the model
reduces to a standard picture.

We demonstrated that, both for slowly running dilaton phase and
constant dilaton phase of the standard scenario, the model can
lead to a small deviation from the $w=-1$ prediction of
non-evolving dark energy; some of the branches, specifically when
the parameter $\delta$ in the one-loop effective action is
negative (or $\xi(\sigma)<0$) can support a non-standard (phantom)
cosmology ($w<-1$), while the other branches support a standard
cosmology with no-acceleration ($w=-1/3$) or acceleration ($-1\le
w < -1/3$) or deceleration ($w>-1/3$).

In our analysis, we have assumed that there is no potential for
the dilaton in the loop-corrected perturbative string effective
action. However, it is quite possible that for the formulation of
a complete and realistic string cosmology scenario, one needs to
take into account the quantum back reactions of loops and
radiation as well as an appropriate non-perturbative dilaton
potential.

We assumed that in our model the contributions arising from two
(and higher) loop terms are suppressed as compared to those from
one-loop terms. In this sense, the result discussed above may
persist in the full theory under certain assumptions about the
moduli dependence of loop corrections to higher derivative terms.
A key point is that a field dependent (and thus space-time
dependent) GB coupling, which is completely reasonable in the
four-dimensional effective action, can make the observability of
various cosmological parameters, including the Hubble expansion
rate of the universe, more achievable.

A more detailed analysis of the model, mainly by introducing
matter fields, is needed yet, first indications are that a GB-type
modification of gravity is a healthy proposal for explaining the
cosmic acceleration problem.

\section*{Acknowledgements}

I would like to thank M. Sami, Shin'ichi Nojiri, Shinji Tsujikawa
and David Wiltshire for helpful discussions and correspondences.
This work was supported in part by the Marsden fund of the Royal
Society of New Zealand, and by the New Zealand Science and
Technology Foundation, under research Grant No. E5229. I also
thank the CERN Theory Division for their hospitality while some of
the work was carried out.

\medskip
\noindent {\it Note Added}: {\small After having sent an earlier
version of this paper to the Journal we learned through
~\cite{Tsujikawa06c} that, for a single scalar field model, a
transition from the matter-dominated era to an accelerating era is
possible for the potential $V=V_0\,\e^{-\lambda \sigma}$ and the
coupling $\xi(\sigma)=\xi_0\,\e^{\mu \sigma}$ when $\mu
>\lambda$. A recent paper by Koivisto and
Mota~\cite{Mota06a} puts various cosmological and astrophysical
constraints for the above choice of $V(\sigma)$ and
$\xi(\sigma)$.}

\section*{Appendix: Field equations as a system of differential equations}
\renewcommand{\theequation}{A.\arabic{equation}}
\setcounter{equation}{0}

In order to express equations (\ref{GB1})-(\ref{GB4}) as a set of
differential equations, we define the following variables:
\begin{eqnarray}\label{def-basic}
&& x\equiv \frac{\dot{\sigma}}{H}, \quad \qquad y\equiv
\frac{V}{H^2}, \quad \qquad
z\equiv \frac{\dot\phi}{H},\nonumber\\
&& u\equiv \frac{\partial \xi}{\partial\sigma} H^2, \qquad
v=\frac{\partial \lambda}{\partial\phi}H^2, \qquad
\varepsilon=\frac{\dot{H}}{H^2}.
\end{eqnarray}
One also notes that $f\equiv \lambda(\phi)-\xi(\sigma)$, and
\begin{equation}
\dot{\lambda}=\frac{\partial\lambda}{\partial\phi}\,\dot{\phi},
\qquad  \ddot{\lambda}=\dot{\phi}^2 \frac{\dd^2
\lambda}{\dd\phi^2} + \frac{\dd \lambda}{\dd\phi}\,\ddot{\phi},
\end{equation}
and a similar expression for $\ddot{\xi}$. A simple calculation
yields
\begin{eqnarray}\label{more-relations}
&& \frac{\ddot\sigma}{H^2} = x^\prime+x \varepsilon,\qquad
\frac{\ddot\phi}{H^2} = z^\prime+ z\varepsilon, \nonumber \\
&&\frac{\partial^2 \xi}{\partial\sigma^2} H^2 =
\frac{u^\prime-2\varepsilon u}{x}, \qquad \frac{\partial^2
\lambda}{\partial\phi^2} H^2 = \frac{v^\prime-2\varepsilon
v}{z},\nonumber \\
&& \frac{\partial V}{\partial\sigma}\frac{1}{H^2}=
\frac{y^\prime+2\varepsilon y}{x},
\end{eqnarray}
where the prime denotes a derivative with respect to $N$, where
$N\equiv \int H \dd t=\ln (a(t)/a_0)$. Note that
$\frac{\partial}{\partial t}=\frac{\partial N}{\partial
t}\frac{\partial}{\partial N}=H\,\frac{\partial}{\partial N}$. The
equations of motion, equations~(\ref{GB1})-(\ref{GB4}), then take
the form
\begin{eqnarray}
&\fl 0= -3+\frac{\kappa^2}{2}\left(\gamma x^2+\zeta z^2\right) +
\kappa^2 y+ 3 \kappa^2 (x u- z v),\label{A4}\\
&\fl 0= 2\varepsilon+3+\kappa^2\frac{\dd}{\dd N}(z v- x
u)+\kappa^2 (\varepsilon+2)(z v- x u)
+\frac{\kappa^2}{2}\left(\gamma x^2 +\zeta
z^2\right)-\kappa^2 y,\label{A5}\\
&\fl 0= \frac{\kappa^2}{2}\frac{\dd }{\dd N} \left(\gamma
x^2\right)+\kappa^2(\varepsilon+3)\gamma x^2 +\kappa^2
(y^\prime+2\varepsilon y) + 3\kappa^2 (1+\varepsilon) x u,\label{A6}\\
&\fl 0= \frac{\kappa^2}{2}\frac{\dd }{\dd N} \left(\zeta
z^2\right)+\kappa^2(\varepsilon+3)\gamma z^2 - 3\kappa^2
(1+\varepsilon) z v.\label{A7}
\end{eqnarray}
Finally, by defining the new variables
\begin{equation}\label{new-variables}
\fl {\mathcal X}=\frac{\kappa^2}{2}\gamma x^2, \qquad {\mathcal
Y}=\kappa^2 y, \qquad {\mathcal W}=\frac{\kappa^2}{2}\gamma
x^2,\qquad {\mathcal U}= \kappa^2 x u,\qquad {\mathcal V}=
\kappa^2 z v,
\end{equation}
we arrive at equations~(\ref{eq1a})-(\ref{eq4a}).

Alternatively, equations (\ref{GB1})-(\ref{GB4}) may be expressed
as a set of first- and second-order differential equations. To
this end, one modifies the definitions for $u$ and $v$ in
equation~(\ref{def-basic}), namely
\begin{equation}
\bar{u}= \xi(\sigma)H^2, \quad \bar{v}=\lambda(\phi) H^2.
\end{equation}
A simple calculation yields
\begin{eqnarray}
&\fl \frac{\partial\xi}{\partial\sigma}\,
H^2=\frac{\bar{u}^\prime-2\varepsilon\bar{u}}{x},\qquad
\frac{\partial^2\xi}{\partial^2\sigma}\,H^2 =
\frac{\bar{u}^{\prime\prime}-2\varepsilon^\prime
\bar{u}-2\varepsilon\bar{u}^\prime}{x^2}-(x^\prime+2\varepsilon
x)\frac{\bar{u}^\prime-2\varepsilon \bar{u}}{x^3},\nonumber \\ \\
&\fl\frac{\partial\lambda}{\partial\phi}\,
H^2=\frac{\bar{v}^\prime-2\varepsilon\bar{v}}{z},\qquad
\frac{\partial^2\lambda}{\partial^2\phi}\,H^2
=\frac{\bar{v}^{\prime\prime}-2\varepsilon^\prime
\bar{v}-2\varepsilon\bar{v}^\prime}{z^2}-(z^\prime+2\varepsilon
z)\frac{\bar{v}^\prime-2\varepsilon \bar{v}}{z^3}.\nonumber
\end{eqnarray}
By substituting these expressions, along with the first, second
and fifth relations defined in (\ref{more-relations}), back to
equations (\ref{GB1})-(\ref{GB4}), and again defining new
variables as in (\ref{new-variables}), except that $\bar{\mathcal
U}=\kappa^2 \bar{u}$ and $\bar{\mathcal V}=\kappa^2 \bar{v}$, we
arrive at equations~(\ref{main1a})-(\ref{main4a}).

Finally, we find several fixed points for the system
(\ref{A4})-(\ref{A7}), characterized by $x^\prime=y^\prime
=z^\prime=0$, and $u^\prime=0$ or $v^\prime=0$.
The fixed-point solutions are specific to the choice of the
potential or the couplings $\xi(\sigma)$, $\lambda(\phi)$, other
than the coupling constants $\gamma$ and $\zeta$; below we will
make a canonical choice, that is, $\zeta\equiv 2\gamma/3$.

For the system of equations (\ref{A4})-(\ref{A7}), there exists a
(global) fixed-point, as given by $(x,y,z)=(0,3,0)$, which
corresponds to a stable de Sitter phase with $\varepsilon=0$. In
particular, in the case of a vanishing potential, so
$y=0$, the fixed-point solutions are given by \\
$\bullet$ (A) $(x, z, u)=(x_0, 0, 1/x_0-\gamma x_0/6)$, with an
arbitrary $v$, and $\varepsilon=-(6+5\gamma x_0^2)/(6+\gamma
x_0^2)$, which
gives a (stable) de Sitter solution only if $\gamma < 0$.\\
$\bullet$ (B) $(x, z, u)=(\frac{-3u_0\pm \sqrt{6\gamma+9
u_0^2}}{\gamma}, 0, u_0)$, with an arbitrary $v$ and
$\varepsilon=-3-\frac{2 u_0\sqrt{3}}{\sqrt{2\gamma+3 u_0^2}}$,
which
gives a de Sitter solution for $u_0 < -\sqrt{2\gamma/3}$.\\
$\bullet$ (C) $(x, z, v)=(0, z_0, \frac{\gamma
z_0}{9}-\frac{1}{z_0},)$, with an arbitrary $u$ and
$\varepsilon=-\frac{9+8\gamma z_0^2}{9+2\gamma z_0^2}$. This
branch
gives an accelerating solution only if $\gamma<0$.\\
$\bullet$ (D) $(x, z, v)=(0, \frac{9v_0\pm
\sqrt{9v_0^2+4\gamma}}{2v_0^2+\gamma}, v_0)$, with an arbitrary
$u$ and $\varepsilon=\frac{-5 v_0^2-3\gamma\pm
v_0\sqrt{9v_0^2+4\gamma}}{2 v_0^2+\gamma}$. Depending upon the
sign of $v_0$, one of the solutions gives a stable de
Sitter phase.\\

If $y\ne 0$, then, for the purpose of finding fixed point
solutions, one needs to specify the functional form of $V(\sigma)$
and/or the coupling $\xi(\sigma)$ or $\lambda(\phi)$. As a simple
case, consider $V=V_0\,\e^{-\,\lambda\sigma}$, which implies that
$y^\prime+(2\varepsilon+\lambda x) y=0$. The fixed point or
critical solutions are then given by \\
$\bullet$ (E) $x=0, \quad y=3+\frac{2\gamma z^2}{3},\quad
\varepsilon=\frac{6\gamma z^2}{9+2\gamma z^2},
\quad v=\frac{\gamma z}{3}$, \\
where $z=z_0\,\e^{-3N}$ and $u\equiv u(N)$. \\
$\bullet$ (F) $ x=\frac{2}{\lambda}, \quad
y=\frac{4\gamma}{\lambda^2}, \quad z=0, \quad \varepsilon =-1,
\quad
u=\frac{\lambda^2-2\gamma}{2\lambda}$,\\
where $v=v(N)$.\\
$\bullet$ (G) $ x=\frac{2}{\lambda}, \quad
y=\frac{4\gamma}{\lambda^2}, \quad \varepsilon =-1, \quad
u=\frac{\lambda^2-2\gamma}{2\lambda}-\tilde{u}$,\\
where $z=z_0\,\e^{-\,3N}$, $\tilde{u}=\frac{\lambda
c}{2}\left(v-\frac{c\gamma}{9}\,e^{-\,3N}\right)$, with $c$ being
an integration constant, and $v=v(N)$.

If one restricts the coupling $\xi(\sigma)$ to be
$\xi(\sigma)=\xi_0\,\e^{\mu \sigma}$, or equivalently
$u^\prime+2(1-\mu/\lambda)u=0$, then the above critical solutions
are further constrained; we now have \\
$\bullet ~~ u=\frac{u_0}{9+2\gamma z^2}$\\
$\bullet ~~ \lambda=\mu $ \\
$\bullet ~~ u=u_0\,\e^{2(\mu/\lambda-1)N}$, \\
respectively, for the branches $E$, $F$ and $G$. It is obvious
that the choice $\lambda=\mu$ is special.

For $\zeta=2\gamma/3$, the system of equations
(\ref{A4})-(\ref{A7}) can be reduced to the following two sets of
equations:
\begin{eqnarray}\label{mono-branch1}
&& \frac{dx}{dN} = - \frac{1}{u} \frac{d
u}{dN}+\frac{2\varepsilon+\gamma
x^2}{u}+x(1-\varepsilon), \nonumber \\
&& y-3+\frac{\gamma}{2} x^2 + 3 u x =0, \quad z=0,
\end{eqnarray}
and
\begin{eqnarray}
&& \frac{dz}{dN}=\frac{6y-18+3\gamma x^2-16\gamma z^2+18
ux}{4\gamma z}=\frac{9(v-\gamma z)}{2\gamma},\nonumber\\
&& \frac{dx}{dN} = x+\frac{\gamma
x^2}{u}+\frac{z}{u}\frac{dv}{dN}-\frac{x}{u}
\frac{du}{dN}+\frac{7\gamma z}{9u} \frac{dz}{dN},\quad \varepsilon=0.\nonumber \\
\end{eqnarray}
For $V=V_0\,\e^{-\,\lambda\sigma}$, from the branch
(\ref{mono-branch1}), we obtain
\begin{equation}
u= \frac{\sqrt{4\gamma^2- 6\gamma\lambda^2}}{3\lambda}\,
\tanh\frac{\sqrt{4\gamma^2-6\gamma \lambda^2}}{2\gamma} (N+N_0)
-\frac{2\gamma}{3\lambda}
\end{equation}
and $x=- 3 u/\gamma$. In addition, if we choose
$\xi=\xi_0\,\e^{\lambda\sigma}$, then we require
$\lambda^2=2\gamma/3$ and hence $du/dN=dx/dN=0$. The stability
analysis of the critical solution $a\propto (t+t_1)^\alpha$ and
$\sigma\propto \ln (t+t_1)$ would be similar to that in
\cite{Nojiri05b}. Further implications of the model will appear
elsewhere.

\medskip
\section*{References}

\end{document}